# A correlation between energy gap, critical current density and relaxation of a superconductor

Harald Reiss[#]

*Department of Physics*
*University of Wuerzburg, Am Hubland, D-97074 Wuerzburg, FRG*
harald.reiss@physik.uni-wuerzburg.de

Abstract

Superconductors like other solids cannot relax instantaneously from excited states to thermodynamic equilibrium. In this paper, relaxation from thermal excitations is investigated, like after absorption of radiation or, under conductor movement, release and transformation of mechanical tension to thermal energy. Relaxation proceeds within finite periods of time the length of which increases the more strongly the closer the superconductor temperature has already approached its critical value. Properties of many-particle systems (as explained, by an analogy to nuclear physics), basic thermodynamic considerations (temperature uniquely defined under solely equilibrium condition) and standard, multi-component heat transfer principles (solid conduction plus radiation in thin films) are applied as tools to prove this expectation. Energy gap, superconductor critical current density and critical temperature, as a result, are tightly related to relaxation rates and relaxation times of the superconductor electron system. By numerical simulations, an attempt is made to find a quantitative correlation of these properties in a thin film superconductor.

## 1   Overview: Superconductor critical parameters and relaxation

This paper presents an attempt to correlate superconductor critical current density and critical temperature with relaxation of its electron system. Focus is on relaxation from thermally excited states. These may result from absorption of radiation or, under conductor movement, release and transformation of mechanical to thermal energy. The reported numerical investigations shall describe how long it takes a disturbed superconductor to relax to thermodynamic equilibrium.

---

[#] Associate (Außerplanmäßiger) Professor

Critical current density becomes zero if, under such disturbances, density of charge carriers (electron pairs) becomes too small to support zero-loss current transport. The relaxation (re-organisation) rate of the disturbed electron system might not be able to compensate decay of electron pairs to single electrons. Accordingly, critical current density should be correlated with relaxation rate and relaxation time of the electron system.

The same applies to critical temperature. As will be shown, if relaxation time after a disturbance is too long, or even might diverge, critical temperature cannot uniquely be defined within reasonable experimental or simulation periods of time.

The literature reports a large number of magnetic relaxation measurements to determine pinning potentials in $Nb_3Sn$, in the BSCCO family and in $MgB_2$/Fe superconductors. Instead, focus in the present paper is on decay of the excited electron system. Relaxation of this system becomes most interesting when, under a thermal disturbance, the superconductor has already approached very closely its superconducting to normal conducting phase transition. Superconductor stability against quench thus is an aspect of the relaxation problem.

In the first, formal part of this paper, therefore, superconductor temperature excursion under a disturbance is revisited in more detail. The presented investigations do not duplicate previous papers [1 - 3]; in particular, superconductor stability is not in the foreground of these investigations. The simulations instead shall safely identify uncertainties of results that might touch or hide the possibly existing correlation between critical parameters and superconductor relaxation.

In the second part, justification of the said, tentative correlation shall be confirmed by quantitative results.

In comparison to numerous studies of magnetization, its relaxation and the pinning potentials, it appears *this* discussion, relaxation rate and relaxation time of the disturbed electron system, has not thoroughly been analyzed in the physical and engineering literature.

This paper is a new edition of the everlasting, general question: What can we really know about superconductor properties? This paper is focused on one just one, yet strongly important aspect of this problem. For this purpose, the reported investigations are applied to a coated (2G) YBaCuO 123, thin film superconductor (Figure 1), for it is this superconductor the knowledge of its materials and transport properties is superior against other superconductors.

## 2    Temperature excursion under a disturbance

To provide a safe basis for investigating the said, possibly existing correlation, criteria and constraints to obtain reliable numerical results by simulation of temperature excursion in a superconductor have to be provided.

### 2.1 Formal calculation of $J_{Crit}$

Critical current density, $J_{Crit}$, depends on temperature,

$$J_{Crit}(T,B) = J_{Crit}(T)/[B_0 + B(t)] \tag{1a}$$

using

$$J_{Crit}(T) = J_{Crit0} (1 - T/T_{Crit})^n \qquad (1b)$$

In a coil, B(t) in Eq. (1a) describes magnetic induction, like from neighbouring windings ($B_0$ is a constant).

**2.2 Decision on the exponent n**

As far as possible, the exponent n in Eq. (1b) has to be chosen with respect to materials and current transport properties. A decision on the value of the exponent can be made if, when using Eq. (1a,b) in numerical simulations, convergence of the simulated temperature fields has been achieved.

Eq. (1b) is derived from Ginzburg-Landau (GL) theory, as reported in Eq. (4-35) of [4]. This result follows from $\partial J_s/\partial v_s = 0$, to find the maximum of $J_s$, a super-current (and with $v_s$ the velocity of its constituents). In Eq. (4-36) of this reference, $J_s$ is identified as the critical current, $J_{Crit}$.

This result of the Ginzburg-Landau equation for $J_{Crit}$ applies to a continuum by assuming the amplitude of the order parameter is uniform in the sample; in thin films, it cannot vary over thickness. Accordingly, we find its applicability to metals and alloys in Sect. 5.6 of [5], at temperature near critical temperature.

But the Ginzburg-Landau theory does not take into account anisotropic materials and transport properties of type II, hard superconductors and thus might not be applicable in these cases. Yet Clem et al [6] explained that Eq. (1b) was applicable for calculation of critical current density near critical temperature in a granular superconductor (NbN; the authors assume an array of Josephson couplings between grains). More applications of Eq. (1b) were reported in several papers [7] to [9] that describe current transport even in granular or thin film, high-temperature superconductors.

We cannot safely apply Eq. (1b) with inclusion of an exponential damping factor, $J_{Crit}(T,B) = J_{Crit}(T=0,B) (1 - T/T_{Crit}) \exp(-T/T_{Crit})$, as was suggested in [10] for investigation of coated superconductors. While this would fit the architecture of the superconductor investigated in the present paper (Figure 1), application of the exponential term would cause serious convergence problems, and $J_{Crit}(T)$ is reduced too strongly.

The traditional Eq. (1b), without an exponential term, therefore has been applied in the present simulations, using the interval $0.5 \leq n \leq 2$, to look for divergences, and after a variety of test calculations, we finally returned to the Ginzburg-Landau value $n = 3/2$.

## 2.3 Decision on $T_{Crit}$

For calculation of the critical current density, $J_{Crit}(T) = J_{Crit0} (1 - T/T_{Crit})^n$, first the difference $\Delta T = T - T_{Crit}$ between superconductor temperature, $T$, and critical temperature, $T_{Crit}$, has to be determined. Second, $J_{Crit}$ contains $T_{Crit}$ by the ratio $T/T_{Crit}$. In both cases, for $\Delta T$ and for $T/T_{Crit}$, a decision has to be made *which* value of $T_{Crit}$ has to be applied in the simulations. At first sight, this is an unexpected question, but we will show that it has strong impacts on calculated temperature excursions.

As a result of the calculations, the most probable, realistic approach is to apply in $T - T_{Crit}$ and $T/T_{Crit}$ *local* values of both conductor temperature *and* of critical temperature. Instead of the standard $\Delta T = T - T_{Crit0}$ and ratio $T/T_{Crit0}$ (using uniform $T_{Crit0}$ = 92 K), we have applied

$$\Delta T(x,y,t) = T(x,y,t) - T_{Crit}(x,y,t) \tag{2a}$$

$$J_{Crit}[T(x,y,t)] = J_{Crit0}\left[1 - T(x,y,t)/T_{Crit}(x,y,t)\right]^n \tag{2b}$$

provided $T(x,y,t) < T_{Crit}(x,y,t)]$ at co-ordinates (x,y) in the conductor cross section.

This differentiation is not just academic, and the temperature fields should be simulated as exact as possible.

Safety in prediction of temperature excursions with time, of critical current density (and, trivially, also of stability functions and thus on onset of a quench) rely on how exact the sharp condition $T \leq T_{Crit}$ can be verified, which means it relies on the uniqueness by which temperature fields, $T(x,y,t)$ *and* critical temperature, $T_{Crit}(x,y,t)$, both as local values, can be determined. Emphasis in this paper is on "local".

## 2.4 Non-uniformity of temperature resulting from statistical parameter variations

Under disturbances, and in particular in situations close to the superconducting/normal conducting phase transition, there is hardly a uniform distribution of temperature and of critical and transport current in the superconductor cross section or volume, compare the results reported in [1 - 3].

Non-uniformity is expected from a variety of potential, materials and physical origins. First, there are materials deficiencies resulting from materials preparation (deposition, handling, conductor winding), application and operation of a superconductor device. Non-local materials properties on a microscopic scale go back to local deviations from perfect stoichiometry, inclusion of foreign atoms, weak links in the grained BSCCO superconductor family, voids, cracks, mechanical tensions arising from conductor movement under Lorentz forces. There is also non-local, magnetic induction induced by currents in neighbouring thin films or filaments, and there are non-local heat transfer mechanisms like scattering of radiation in thin films and non-local solid/liquid interactions at the interface to the coolant.

Trivially, as far as the set of available *simulation parameters* is concerned, non-uniformity of the temperature fields and excursions with time may arise also from experimental errors during measurements of $T_{Crit}$ and of the other critical variables of the superconductor.

The simulations of the coated, (2G) YBaCuO 123 thin film superconductor like in previous paper apply statistical variations $\Delta T_{Crit}$, $\Delta J_{Crit}$ and $\Delta B_{Crit}$ (the upper critical magnetic field). In the present paper, as an additional aspect, also a statistical uncertainty of the thermal diffusivity of the superconductor material is considered.

As an optimistic, provisional estimate, a spacing $\Delta T_{Crit} < \pm 1$ K and likewise variations $\Delta J_{Crit0}$ and $\Delta B_{Crit20}$ are assumed as within 1 percent and $\pm 5$ Tesla, respectively, in the simulations. The variations $\Delta T_{Crit}$, $\Delta J_{Crit0}$ within these given, maximum uncertainty intervals are shown in Figure 2. In reality, uncertainties of $J_{Crit}$ and $B_{Crit}$ may be much larger than the assumed $\Delta J_{Crit} = \pm 1$ percent and $\Delta B_{Crit20} = \pm 5$ T; the calculations presented here shall demonstrate the principal impact of these uncertainties on calculated results, see later, Figure 5b.

Though experimental methods meanwhile have significantly been refined by increasingly sharp detection criteria (like electrical field constraints below $10^{-6}$ V/cm in transport measurements of $T_{Crit}$), the extension by which experimental results really deviate from expected (true) values has remained an open question. What we mostly find in the literature are error bars resulting from estimates applied to final experimental results. A statistical approach applied in the most early stage of the simulations, like the one presented in the present paper, is a better solution.

Random variations of the thermal conductivity, within ± 5 per cent and first results showing their impact on centroid element temperature are described in the Appendix, Figure 10 (it would be extremely difficult to achieve experimental uncertainty of less than five percent in the conductivity or thermal diffusivity of thin films).

Simulations using non-uniform superconductor critical parameters is contrary to all traditional, numerical calculations of temperature fields under disturbances (and also to all standard stability calculations).

But the most interesting question is whether critical temperature of a superconductor can uniquely be defined at all. Existence of a uniquely defined $T_{Crit}$ depends on the thermodynamic state of the superconductor electron system[1]. Trivially, any uncertainty of $T_{Crit}$ would concern the differences $\Delta T(x,y,t)$ and the ratios $T(x,y,t)/T_{Crit}(x,y,t)]$ when calculating $J_{Crit}[T(x,y,t)]$ by Eq. (2b).

---

[1] Critical temperature like critical current density is a property of solely the electron system. It is an interesting question which temperature is measured, electron or phonon temperature, when sensors attached to a superconductor solid or thin film, or remotely by radiation detectors. It is not clear that both temperatures would be identical, compare Sect. 4.3 in [11] for results obtained with different superconductors.

## 2.5 Uncertainty of $T_{Crit}$ from relaxation time

Relaxation of the superconductor electron system from a disturbance is expected to follow an exponential decay law.

Let ς(t) denote an excitation like a sudden temperature increase following e. g. absorption of a radiative pulse. The excitation disturbs the dynamic equilibrium between statistical but permanent, single electron (quasi-particle) generation and their recombination to electron pairs. The excitation decays proportional to exp(- t/т), with т the relaxation time, traditionally assumed as constant. But it is not clear that т is really constant; the density, $n_S(T)$, of electron pairs strongly depends on temperature that in turn, under a disturbance, is a function of time, t.

The point is: Relaxation cannot be completed instantaneously. Trivially, relaxation time is correlated with relaxation rates. Spontaneously, one would assume that relaxation time increases the stronger, the smaller the relaxation rate (the longer it takes the superconductor to relax to a new thermodynamic equilibrium). Regardless how relaxation rates are calculated, like in [11], see below, or from Eq. (8) of [12], the rates converge to zero when temperature approaches critical temperature of the superconductor.

An approach how to calculate relaxation time, т, and its dependence on temperature has been suggested by the recently reported, "microscopic stability model" [11] Its details shall not be repeated here, a short description can be found in [17] when the model was applied to a (2G) thin film superconductor. But its consequences have to be taken into account when inspecting the said relaxation problem.

In short, the model [11] is a multi-physics approach using four formal analogues to describe the *temporal* aspect of the re-organisation of the total electron system after a disturbance. Relaxation not only means recombination of a limited number of single electrons (the decay products from a disturbance) to electron pairs but re-organisation of the *total* electron body, i. e. all electrons in dynamic equilibrium with electron pairs of the superconductor. The method accordingly includes

(a) Calculation of "coefficients of fractional parentage" (cfp's), a concept applied in atomic and nuclear physics; here, this concept describes the total number of electron states involved in the re-organisation of the total electron body and, after completion of the relaxation, allows to calculate the total time needed to arrive at the new, thermodynamic equilibrium,
(b) a "time of flight"-concept with a mediating Boson. In nuclear physics, the Yukawa-model, as one of possible analogues, a pion, $\pi$, mediates coupling (binding) of two nucleons, like the n and p in a Deuteron.
(c) the Pauli selection rule, and (d) the uncertainty principle.

In comparison to this model, derivation of the superconducting electron density, $n_S(T)/n_0 = 1 - (T/T_{Crit})^4$, Eq. 8 of [12], with $n_0$ the total number of electrons at T = 0, from Eqs. (2) to (7) of the same reference, neglects the dynamic aspects (statistical, quasi-particle generation and corresponding recombination processes) of the problem. Yet comparison of the curves "relaxation rate vs. temperature" obtained from both [11] and [12] shows at least qualitative agreement (yellow and green diamonds in the Appendix, Figure 12a).Differences result just from assumptions on numerical values of electron density, $n_S(T)$.

Both the direct, multi-step method described in [11] and, with some additional assumptions, the possibly heuristic Eq. (8) of [12], not only allow calculation of relaxation rates and relaxation time. A far-reaching conclusion from relaxation time obtained in [11] (here reproduced with Figure 12b, again in the Appendix of the present paper) can also be drawn about existence of critical temperature, to be understood as a unambiguously defined, thermodynamic quantity.

If $T_{Crit}$ exists, it can be understood as being either
- a rough description of a thermodynamic *non-equilibrium electron* state (this is a standard but only approximate interpretation), which means, from purely thermodynamic standpoints, $T_{Crit}$ cannot be explained as a thermodynamic temperature of uniquely defined value,
- or $T_{Crit}$ correlates, as a thermodynamic *equilibrium* variable, with a corresponding, completed (which means, really existing) equilibrium electron state. The prediction from [11] (and could be drawn also from Eq. (8) in [12]) is that in this case the new equilibrium electron state cannot be reached within finite process or simulation time.

The situation is schematically described in Figure 3. Relaxation time increases the more the closer the electron system approaches the phase transition because the more electrons have to be re-organised to electron pairs to obtain the new equilibrium.

This conclusion is not in conflict with Buckel and Kleiner [13], Chap. 4, p. 262: The authors say that "in conventional superconductors the concentration of unpaired electrons decreases exponentially with decreasing temperature, and hence the probability that an unpaired electron finds a suitable partner for recombination to form a Cooper pair also decreases." This is not in conflict with the prediction from [11] because the situation described in [13] is strongly different.

In the present case, under increasing temperature, the number of single electrons increases (quasi-particles as thermally, not permanent, statistically initiated decay products). Starting from an original dynamic equilibrium state at a temperature, T, followed by a disturbance at *this* temperature, increasingly *more* single particles have to recombined to pairs in order to generate, by reorganisation of the total number, $N_{Total}$, the new dynamic equilibrium, at the *new* equilibrium temperature, T'. There, again a statistical, un-displaced dynamic equilibrium between generation and recombination processes would be obtained when relaxation is completed. Conservation of energy requests T' > T; this temperature increase, from T to T', concerns *all* electrons (the total electron body). The increased number, $N_{Total}(T')$, of electrons to be reorganised accordingly *increases* relaxation time (while, reversely, relaxation time decreases with decreasing temperature). It is the fractional parentage principle that causes the enormous increase of τ observed near $T_{Crit}$ in Figure 12b.

Instead, the explanation [13] and the original papers by Gray et al. [14, 15] do not refer to a temperature increase from absorption of a radiation pulse or from other thermal disturbances that would affect the total body of electrons. Contrary to the situation described in [11], they describe injection experiments to create an additional number ΔN of quasi-particles above the "core" (N) of bound electrons in the undisturbed superconductor. The excursion with time of the number ΔN, the relaxation time, τ, is described by equations of the type $\partial \Delta N / \partial t = A + B$, with sources, A, and sinks, B. The contribution A contains the injection rate, B comprises different types of quasi-particle losses resulting from e. g. escape processes and by interaction with lattice phonons. These equations do not take into account the whole electron body (N) but only the additional, injected number ΔN of particles. It is thus not the total body of electrons that has to be re-organised in this model under observation of the fractional parentage principle.

The relaxation time reported in [14, 15] therefore is strongly different from the predictions made in [11]. The model [11] does not contradict but simply cannot be applied to the injection experiments [14, 15], and, reversely, the relaxation time obtained from quasi-particle injection has little relevance for relaxation of from thermal excitation.

As a consequence, even if it in a warm-up experiment the temperature of the electron system might become infinitely close to the traditional $T_{Crit}$, re-organisation of the corresponding, total electron state still is not completed and thus cannot be understood as a thermodynamic equilibrium. Critical temperature, as a thermodynamic equilibrium quantity, accordingly is not uniquely defined.

If temperature evolution in a superconductor is not uniform, the phase transition at critical temperature can be completed neither spatial uniformly nor will this final state be attained simultaneously in the superconductor cross section or volume.

The consequence to be taken from [11] concerning non-existence of a uniquely defined $T_{Crit}$ is absolutely contrary to standard understanding of critical temperature (traditionally, $T_{Crit}$ is, so to speak, under "monumental protection"). Instead, electron states near the phase transition, and thus near (traditional) critical temperature, are hidden from direct experimental observation and from the results obtainable in simulations.

Critical temperature, if understood as a sharply defined quantity, therefore is a fiction, it remains out of sight during finite (experimental or simulated) periods of time. Rather, it is not *at $T_{Crit}$* that simulations can be performed but in a time interval *within which* events and their distance from the phase transition can be studied.

The conclusion from [11] also specifies a note that Annett [16], p. 52, added to the left column of his book. There we have "The word superconductor is used only to mean a material with a definite phase transition and critical temperature." (correction by the present author: "definite" probably means "completed"). Yes, but the phase transition, here the completion of relaxation after a disturbance, takes finite, possibly diverging time. Critical temperature is the dynamic equilibrium of the electron state when *no more disturbances*, besides the *statistical* fluctuations in dynamic equilibrium, have to be compensated. Critical temperature thus may become out of sight within reasonable time scales in standard experiments or in computer simulations.

This conclusion is not new physics but results from properties of many-particle systems (an analogy to nuclear physics), basic thermodynamic considerations (temperature uniquely defined under thermodynamic equilibrium) and from standard, multi-component heat transfer principles (solid conduction plus radiation in thin films).

Consequences from this conclusion not only concern experiments and simulations but might concern also levitation: Levitation height might not converge within reasonable periods of time to a final, equilibrium vertical co-ordinate. Even more important, but obviously inevitable under reasonably limited experimental efforts, results of standard measurements of critical current density might not reflect perfect equilibrium conditions.

All results reported in the following Figures therefore are provisional in the sense that, strictly speaking, simulations can safely be performed *only if* a *convergence limes* of a series $T_{Crit}$ really would exist.

## 3   Simulation Scheme

The following relations between critical current density and relaxation time, τ, or relaxation rate, ζ,

$dJ_{Crit}/d\tau \sim J_{Crit}$, or $\quad\quad\quad\quad\quad\quad\quad\quad\quad\quad\quad\quad\quad\quad$ (3a)
$dJ_{Crit}/d\zeta \sim J_{Crit}$ $\quad\quad\quad\quad\quad\quad\quad\quad\quad\quad\quad\quad\quad\quad\quad$ (3b)

are strongly non-linear, solutions can be found only in numerical simulations, and they can be quantified only if $J_{Crit}$, τ and ζ depend on a common variable, which here is temperature. The simulated temperature excursions with time shall contribute to quantitative solutions of Eq. (3a,b).

If inspected under the item "simulation time", the overall procedure yielding relaxation rates and relaxation time in dependence of (and correlated with) critical current density is schematically explained in Figure 3.

## 3.1  Overall procedure

We start with simulations of transient temperature fields, T(x,y,t), developing from initial values, T(x,y,$t_0$), in a superconductor coil. The coil integrates 100 turns prepared from a coated, (2G) YBaCuO 123 thin film superconductor (Figure 1). Details of conductor architecture are described in [17] and will not be repeated here. The transient temperature distributions under disturbances in this paper are obtained from 2D solutions of Fourier's differential equation obtained by means of the Finite Element (FE) method using a standard, commercially available FE code.

This start of the simulations is chosen because temperature, like magnetic field, is a thermodynamic variable that deserves primary attention, while current is not. We do not apply a multi-physics approach like H- or T/A formulations that beyond doubt are suitable for large-scale superconductor applications, see [18] and references cited therein, in particular the review by Shen B, Grilli F, and Coombs T (August 2019).

Instead, what is needed in the present paper, with respect to the said, possibly existing correlation, requests electromagnetic/thermal coupling without homogenization but with modelling the physics on *microscopic* scales. We have to consider, in each of the Finite Elements, the Meissner effect, weak links and viscous, thermally assisted flux flow resistivity and flux creep if the superconductor is very close to critical temperature and contact resistances between super- and normal conducting layers - how else could non-uniformity of the temperature distributions be considered. We do not apply constant temperature approximations but describe the *local* interplay of temperature-, field- and current-dependent variables with the transient temperature fields.

Standard Finite Element codes provide multi-physics element options (like the electric field, heat transfer, structural mechanics couplings by the FE code Ansys). But it is not clear that convergence can safely be achieved if the superconducting/normal phase transition, with its complicated consequences for the other superconductor variables, would be involved in the solutions obtained from standard FE codes.

Therefore, we first calculate the solution, T(x,y,t), by application of the FE code to highly differentiated, geometrical and materials boundary conditions. From *this* solution, all physical variables that depend on temperature are calculated. In successive load-steps, the Finite Element integration is restarted with the new variables; Bi-directional interconnection between electric/magnetic and thermal interplay is realised by this sequential procedure. The convergence problem is reduced to that of solely the temperature field, yet the procedure requires considerable computational efforts.

The FE procedure for this purpose is embedded in a comprehensive, "master scheme". This scheme, a flow chart that is extended against the previously published Figure 12 of [19], is shown in the Appendix of the present paper, Figure 11a). Step by step repetitions of the proper Finite Element integration cycles are intersected by intermediate, analytic calculations of electrical and magnetic variables (penetration of magnetic field, Meissner effect, local resistances, local critical and transport current distribution and local losses like flux flow, Ohmic or inductive).

The FE procedure during its many repetitions allows to introduce corrections (adjustments) to physically reasonable values like the 77 K

coolant temperature constraint in case of divergences. This is realised without restarts of the simulations.

## 3.2 Selection of the number of simulated turns

Another serious problem arises from the number of turns that can be simulated to yield high spatial and temporal resolution, without application of any continuous, spatially averaged structures. The number of simulated turns for this purpose has to be limited.

In the present simulations, the procedure has been applied to only the five uppermost turns of the coil. Accordingly, a thermal boundary condition near the interface between turns 96 and 95 has to be specified appropriately.

If $y_{96}$ denotes vertical position at half the thickness of the Hastelloy block, the condition $T(y_{96}) \equiv 77$ K (temperature of the coolant) would not be realistic because temperature of turn 95 is much larger, at least in case of a fault current. Any values $T(y_{96}) > 77$ K at this position could only be estimated, but estimates shall be avoided.

In order to solve this problem, the plane $y_{96}$ = const in the present calculations has been defined as plane of vertical symmetry, with temperature of turn 96 reflecting temperature of turn 95 which in turn is close to temperature of turns 94 and so on. This is because all turns are tightly, by strong, solid/solid contacts, coupled to each other. Physically, deviations from symmetry can arise only close near the upper boundary of the coil with its solid/liquid contact. At this co-ordinate, temperature of turn 100 *physically* (with respect to thermal transport) does not reflect temperature of turn 92 while it *numerically* does.

This deviation from reality is acceptable because, physically, there are many intermediate layers between turns 92 and 96, and these are expected to thermally insulate turn 96 from the turns below (needless to say, the turns are also electrically insulated against each other).

Numerically, symmetry is specified by leaving the boundary condition open at the co-ordinate $y_{96}$, a measure that is frequently applied in numerical (Finite Differences or Element) calculations.

For the upper five turns (96 to 100) of the coil, the FE model applies between 65 000 and 198 000 Finite Elements (Ansys 16, plane55 with 4 nodes). From these, the proper superconductor thin films in each turn occupy between 5 000 and 15 000 elements.

### 3.3 Compensation of convergence problems

We have compensated 2D interfacial layers (IFL) between the thin SC, Ag and MgO-films (that otherwise might be approximated with shell elements) by assuming regular, 2D thin films of very small thickness.

In the total winding cross sections, film thicknesses are between 40 nm (interfacial layers) and 100 µm (stabilizer). With a width of the windings of about 6 mm, and in particular with respect to details of the previously suggested flux flow cell model [1], the element aspect ratio might become larger than limits set by the Finite Element method, but this was avoided by limiting the horizontal element width by increasing the number of elements in each turn. The cell model [1] simply replaces well-known standard relations for solid state, normal conduction resistivity, $\rho_{NC}$, by a conducting network incorporating weak-links to estimate flux flow resistivity, $\rho_{FF} = \rho_{NC,eff} B/B_{Crit,2}$, from the effective electrical resistivity, $\rho_{NC,eff}$, of this network.

Divergences like T < 77 K, or thermal run-away to extremely high temperature, were observed in only rare situations during the iteration cycles. These were located outside the proper superconductor cross section (compare Figure 11c in the Appendix) and could result from too large a time step, too small an element volume in the numerical procedure or from inappropriate convergence criteria.

Control of stagnation temperature confirmed the FE mesh was correctly designed in the present simulations.

Multiple repetitions of the FE calculation steps, as indicated in Figure 11a,b, are unconventional, but this scheme is justified in view of the non-linearity of almost all input variables and parameters, in particular near the phase transition. Against standard, Finite Differences or multi-physics Finite Element procedures, control of convergence by this concept seems to be strongly improved.

### 3.4 Thermal and radiative parameters

Like in the previous papers [1 - 3], the simulations apply temperature-dependent thermal conductivity and specific heat of all materials, see [17] (manuscript and Table 1); these values and the anisotropy of YBaCuO 123 material are based on experiments, either reported in the literature or result from co-operations of the present author with J. Fricke's group in experiments performed at Wuerzburg University.

Radiative transfer in thin films has been modelled as a conduction process since the films are non-transparent to mid-IR radiation [20]. The extinction and scattering properties have been calculated from application of rigorous scattering theory and of the Drude-Lorentz model to estimate the complex refractive indices as a function of temperature, see the results obtained in the mid-IR spectral range for refractive index, extinction coefficients and Albedo shown in Figures 12 to 15 of [20].

More details of the general simulation procedure (meshing, integration time steps, heat transfer to coolant, depairing current density) have been reported in Sects. 2 and 3 of [21], there for the (1G) BSCCO 2223 multi-filamentary superconductor.

## 4  Transient temperature excursions

Figure 4a confirms previous results that temperature distribution within the thin superconductor film is not uniform, with approximately a linear temperature gradient within the 2µm thin YBaCuO 123 film. Figure 4a is a detail of the present results and confirms non-uniform temperature distributions like those reported in previous papers [1 - 3].

Non-uniformity of the temperature distribution is also confirmed by the standard deviations from mean value of the curves in Figure 4b (turns 96 and 100 of the coil). In the outer turns (here turn 100, lower diagram), almost no difference (within their standard deviations) can be detected if calculations are performed with or without the random variations of $T_{Crit}$, $J_{Crit}$ and $B_{Crit1,2}$.

Because of the non-uniform temperature distribution, local zero and Ohmic resistances during finite time intervals exist in parallel, even in turn 100. Also local flux flow resistances, between $77 \leq T(x,y,t) \leq T_{Crit}(x,y,t)$ then exist if local transport exceeds local critical current density at these positions. For $T < T_{Crit}$, it is thus nominal current, flux flow resistance and corresponding losses (and the random fluctuations of the critical parameters) that are responsible for the local increase of $T(x,y,t) > 77$ K, the start value of the simulations, seen in Figure 4b (and later, upper diagram of Figure 11c).

We have applied the ratio of transport to critical current, $I_{Transp}/I_{Crit} = 1$. While this defines the *overall* transport current ratio in the conductor cross section, it does not exclude any local overrun of local critical current density, $J_{Crit}(x,y,t)$, by the transport current density, under its random variations $\Delta J_{Crit}(x,y,t)$.

Since distribution of flux flow and Ohmic resistances may be different at each longitudinal (z) co-ordinate of a thin film or filamentary superconductor, transport current may percolate through length of the conductor. Its distribution at different z-co-ordinates then does not reflect profiles of laminar liquid flow.

Random, local variations of $T_{Crit0}$ and $J_{Crit0}$ obviously are sufficient to induce, without any fault current, just with transport current equal to nominal current, non-uniform temperature distributions. The results, here obtained under parameter sensitivity testing, confirm the findings reported in Figures 4 to 7 of [17]. Without the random variations of $T_{Crit}$, $J_{Crit}$ and $B_{Crit}$, and if density of transport current is below critical current density, the temperature distributions would be flat and would not show any temperature run-away to catastrophic divergence.

By the very large number of elements, trivially the *average* of the random distributions of $J_{Crit0}$, $T_{Crit0}$ and $B_{Crit20}$ almost perfectly coincides with their physical (standard) values. However, the problem is that a certain number of local values of $J_{Crit}$, $T_{Crit}$ and $B_{Crit2}$ possibly might diverge from the uncertainty (percentage) range in Figure 2. A specific value $T_{Crit}(x,y,t) < T(x,y,t)$ then would be responsible for development of a hot spot at the temporal position (x,y,t) even if the stability function, $\Phi(t)$ remains below $\Phi(t) = 1$ (explanation of $\Phi(t)$ and an example is given in Figure 9a,b in the Appendix).

The superconductor literature speaks of quench as an event during which its total magnetic and mechanical tension energy is suddenly released to the total volume. It does not seem inappropriate to apply the term "quench" to also *local,* sudden transformations of magnetic and mechanical to thermal energy: The mechanisms of this transformation is, on a macroscopic scale, the same as their macroscopic counterpart; roughly speaking, they proceed on just different spatial and time scales. But a local quench in many situations may well be the origin of a subsequent, total and catastrophic quench that would cover the total superconductor volume.

Like in [1 - 3, 11, 17, 19 - 21], the calculations are not design calculations. The numerical experiments instead investigate basic superconductor problems using the Finite Element procedure, and like in [3], Monte Carlo simulations, and radiative transfer calculations in thin films, as a tool to explore, as far as possible, the physics behind quench. The investigations rather demonstrate potentials and also limitations of numerical methods in the field of superconductivity. Yet the simulations and the reported results might allow conclusions for materials development, see the second part of this paper.

Computational efforts are enormous. Simulations shown in Figures 4a,b, 5a,b and 11c, over a period of below 5 ms, on a standard PC, with 4-core processor PC and 16 GB workspace under Windows 7, frequently took more than 24 hrs. This mainly results from the large number of IF…THEN…ELSE-steps to be executed separately in each of numerous FE elements, within each under a large number of calculation steps and iteration cycles (Figure 11a,b) and with respect to all potentially open superconductor, zero, flux flow or Ohmic resistance states.

### 4.1 Centroid temperature under parameter variation

In Figure 5a, dependence of the centroid temperature, $T(x,y,t)$, on variations of the exponent n in Eq. (1b) is reported. The observed temporal structures of $T(x,y,t)$ at $t = 4.1$ and $t > 4.2$ ms in the centroid element at first sight indicate onset of a first and of a second (apparent), local temperature run-away. It becomes "real" not before we enter during the iterations the convergence circles indicating the periods (load steps) within which convergence of temperature is achieved. That in both cases the local temperature increase might indicate also a *total* quench, in at least one turn, is confirmed in Figure 5b (lower diagram) and, in the Appendix, in Figure 9a-c by the calculated stability functions and heat fluxes.

The impact of the exponent n in $J_{Crit}(x,y,t) = J_{Crit}(x,y,t_0) [1 - T(x,y,t)/T_{Crit}]^n$ on local (again centroid) temperature is explained by $J_{Crit}(T) = J_{Crit0} (1 - T/T_{Crit})^n$, Eq. (1b), that can be interpreted in two ways:

(1) the larger exponent, n = 1.5, decreases $J_{Crit}$ rather strongly while for all n < 1, $J_{Crit}$ *increases* because [1 - T(x,y,t)/$T_{Crit}$)] < 1 for all T(x,y,t) below $T_{Crit}$. At constant T(x,y,t), and if n = 1.5, superconductor temperature thus becomes more vulnerable to occurrence of flux flow losses. Flux flow losses (if transport current density exceeds $J_{Crit}$) then occur with increasing probability, and T(x,y,t), because of these very losses, becomes larger for n = 1.5, in comparison to n = 0.5.

(2) under given $J_{Crit}(T)$, $J_{Crit0}$ and exponent n, Eq. (1b) after elementary transformations yields

$$T = (1 - [J_{Crit}(T)/J_{Crit0}]^{1/n})T_{Crit} \quad (4)$$

for temperature, T, calculated for different $J_{Crit}(T)$ and exponents n, see Figure 6. For given $J_{Crit}(T)$, temperature T increases if n decreases. This explains the results shown in Figure 5a,b: The smaller the exponent n, the larger the conductor temperature.

This is not in contradiction to item (1) that predicts at constant T, that critical current density $J_{Crit}(T)$ (not temperature, T) increases (and thus flux flow loss decreases) if n decreases.

Figure 5b shows the joint impact of variations $\Delta T_{Crit0}$ and $\Delta J_{Crit0}$ when the random values given in Figure 2 and of $\Delta B_{Crit20}$ are reduced by a divisor (here between 5 and 100) applied to the originally assumed uncertainties of these parameters. In the top turn 100, the mean temperature near $T_{Critt0}$ (lower diagram of Figure 5b) increases by about 1.7 K. Optimistically, this is about the experimental uncertainty of a traditional $T_{Crit}$ measurement itself when it is obtained from transport measurements (with sharp electrical field criteria). In the inner turns, the impact is even larger than in turn 100.

In conclusion from this Section, uncertainties in materials and transport properties of the superconductor cannot be neglected, not in the present simulations (and also not in any stability analysis). Simulations of superconductor transient temperature excursion have to apply

- $\Delta T(x,y,t) = T(x,y,t) - T_{Crit}(x,y,t)$,
- $J_{Crit}[T(x,y,t)] = J_{Crit0} [1 - T(x,y,t)/T_{Crit}(x,y,t)]^n$,

in both cases with *local* $T_{Crit}(x,y,t)$, instead of the simple comparison T vs. $T_{Crit0}$ and instead of uniform $T_{Crit0}$ = 92 K.

More details showing the impact of variations of the exponent n and of $T_{Crit}$ (uniform or spatial resolved) are shown in the Appendix, Figure 9a,b (the impact on the stability function).

It is interesting to note that the (first) steep increase of centroid and mean temperatures of the turns (Figures 4b and 5a,b) at about t = 4.1 ms after start of the disturbance (and correspondingly also of the stability function), all correlate with onset of divergence of the numerical calculation scheme. The same correlation can be seen from the sudden increase of the number of FE equilibrium integrations to obtain convergence when in Figure 1 of [19] the multi-filamentary (1G) superconductor BSCCO 2223 was investigated.

The question now is: *How often* will $J_{Crit}(x,y,t)$, $T_{Crit}(x,y,t)$ and $B_{Crit2}(x,y,t)$, at a given simulation time, really fall into the uncertainty intervals of Figure 2 or into other intervals around the mean values of the critical parameters? The following discussion is focused on $T_{Crit}(x,y,t)$.

## 4.2 Frequency by which $T_{Crit}(x,y,t)$ falls into uncertainty intervals around $T_{Crit0}$

In Figure 7, we control the number of element temperatures falling into deviation intervals δT from critical temperature. Results are calculated for either the deviation δT from $T_{Crit0}$ = 92 K ($T_{Crit0}$ uniform, solid diamonds) or from $T_{Crit}(x,y,t)$, solid triangles (the latter are those of Figure 2, lower diagram). Clear differences are seen only for simulation time t < 4.16 ms. The calculations are performed using the anisotropy ratio, $D_{ab}/D_c$ = 5, of the thermal diffusivity and the exponent n = 1.5 in $J_{Crit}(x,y,t) = J_{Crit}(x,y,t=0)$ $[1 - T(x,y,t)/T_{Crit}]^n$. When instead using n = 0.5, the results are almost identical.

The results show that the statistical frequency distribution of deviations between local element temperature from critical temperature in only rare cases is below 0.01 K. In this case, differentiation would be of little importance for the present simulations since the frequency of these deviation accounts for only 0.14 percent of the total. But deviations of 0.1, 1 and 10 K at t = 4.1 ms amount to about 1.3, 12.3 and 86.3 percent, respectively. The impact exerted by δT = 1 K on the results, by its approximately 12 percent contribution, is already substantial.

# 5 Second part of the paper: Correlation of critical current density with relaxation

## 5.1 The link provided by temperature

The investigation as before is focused on situations near the phase transition. Element temperature serves for calculation of the critical current density (Eq. 1a,b), without field dependence.

The fields $T(x,y,t)$ and $J_{Crit}(x,y,t)$ so far rely on solely results of the Finite Element (FE) simulations, with application of the relation $J_{Crit}(x,y,t) = J_{Crit}(x,y,t = 0) [1 - T(x,y,t)/T_{Crit}]^n$ using n = 1.5 as the exponent and Eq. (2a) for temperature difference ΔT and ratio $T/T_{Crit}$ in Eq. (2b).

In any electrical conductor, the relationship between current density, J, and the density, $n_q$, of involved charge carriers, by conservation of charge, is

$$J = n_q \, v \, q \qquad (5)$$

with v the velocity of the charge carriers and q the moving electric charge. Eq. (5) applies to stationary current flow. It uniquely couples charge density, $n_q$, with current density, J. The density, $n_q$, has to be provided from a model or from experiments.

In superconductors, transport or screening currents all flow with critical current density, $J = J_{Crit}$, at a temperature, T,

$$J_{Crit}(T) = J_{Crit,0} \left( 1 - T/T_{Crit} \right)^n \qquad (6a)$$
$$= n_S(T) \, v_{Fermi} \, 2e \qquad (6b)$$

The temperature fields, T(x,y,t), are mapped onto the field of critical temperatures to yield the fields $J_{Crit}$(x,y,t).

Density of the charge carriers, $n_S$(T), in the superconductor YBaCuO 123 has been calculated in Figure 8a of [19]. We have used Eq. (6a,b) in this Figure to calculate the minimum density of electron pairs (charge carriers), $n_S$(T), necessary to support a given critical current density. The results were safely below the available $n_S$(T) obtained from application of the microscopic stability model [11].

Density, $n_S$(T), of charge carriers and critical current density according to Eq. (6a,b) are coupled to each other. The same applies to their finite variations, $\Delta n_S$(T) and $\Delta J_{Crit}$, both within Eq. (6b), and thus applies to also their ratio $\Delta J_{Crit}$(T)/$\Delta n_S$(T), at given $\Delta J_{Crit}$. In the following, with the decay rate $\zeta = \Delta n_S(T)/\Delta t$, ratios $\Delta J_{Crit}/\Delta \zeta$ of the variations $\Delta J_{Crit}$(T) and $\Delta n_S$(T) shall be calculated from the transient temperature fields obtained in the first part of this paper. The decay rate $\zeta = \Delta n_S(T)/\Delta t$) becomes very small (see Figure 13 in the Appendix).

Figure 8a (blue diamonds in the upper diagram) shows ratio $\Delta J_{Crit}/\Delta \zeta$ of the (finite) $\Delta J_{Crit}$ and $\Delta \zeta$ (not of the variables $J_{Crit}$ and $\zeta$), in dependence of $J_{Crit}$. The ratio in the following is considered as a continuum variable, i. e. as an approximation to $d\Delta J_{Crit}/d\Delta \zeta = f(J_{Crit})$. When this curve is integrated, it yields the variation $dJ_{Crit}$ of $J_{Crit}$ (not the proper $J_{Crit}$ itself) under given relaxation rates, $d\zeta$, and at given $J_{Crit}$, see next Subsection.

Variations of $J_{Crit}$ according to Figure 8a (upper diagram) depend strongly on variations of the relaxation rate: The smaller $J_{Crit}$ (i. e. at temperature close to $T_{Crit}$), the more increases the derivative $d\Delta J_{Crit}/d\Delta \zeta$, and the more sensitive will $J_{Crit}$ react to variations of the decay rate. Obtaining large $d\Delta J_{Crit}/d\Delta\zeta$, therefore, under this specific aspect, could be of practical, possibly also of theoretical interest.

Coupling between $J_{Crit}$ with also materials properties in type I and type II superconductors simply is a matter of fact. A key to increase $J_{Crit}$ thus might be found not only from progress in materials development (like increase of flux pinning, reduction of weak-links) but also if a *correlation* (in its exact sense, see next Subsection) could be identified between relaxation rates and materials properties.

Critical (and transport) current density of thin film superconductors is not, or only weakly, limited by grain boundaries, in contrast to grain structure and corresponding weak link behaviour of the BSCCO superconductor family. What then can be done, from materials transport property aspects, to accelerate relaxation rates?

## 5.2  A correlation between relaxation and materials properties?

Correlation has to be differentiated from causality and from simple coupling.

Coupling is a weak, not necessarily quantifiable relation between two or more objects, events or variables. Correlation indicates strength by which two or more of these are coupled to each other. Coupling might be statistical.

If an event 1 leads to a change of the properties (or simply of co-ordinates) of an event 2, and if this change depends solely on the interaction of event 2 with event 1 (not by others), and if the interaction can be detected and is quantifiable (and possibly reversible like in a resonance), we speak of causality. Causality is stronger than correlation. Correlation trivially includes coupling.

An indicative relation, in the strict mathematical sense, exists between conductor temperature and critical current density of a superconductor. The relation neither is surjective nor is it bijective. Increase of conductor temperature from all experience reduces critical current density, which means both variables are correlated under uni-directional causality. Causality in this case is not bijective (is not reversible).

Consider Figures 12a,b and 13: The larger the conductor temperature (the smaller the remaining difference to critical temperature), the larger are decay rates, and, based on [11], the longer takes it the system to relax to a new thermodynamic equilibrium. Accordingly, temperature, decay rates and relaxation time apparently are correlated, directly or inversely (though again not reversibly). By analogy, can we explain the apparently existing relations described in Figure 8a between critical current density and relaxation as *correlations* and, in particular, as also of *causal* order?

The curves and arrows in Figure 8b, while they suggest that correlations would exist, perhaps could alternatively be explained on just *numerical* grounds; in reality, the correlations could be spurious (or even nonsense correlations). Also calculation of correlation coefficients would not prove physical correlations (and thus causal order) among the variables because these coefficients would have to be based on the same numerical data.

To answer the above question "causality or not?", it has to be checked whether the apparent correlations are uni- or bi-directional and can be explained from the *physics* behind. Can we, from the physics behind, verify that the arrows 1 to 4 in Figure 8b indicate causal relations?

Compare again the blue diamonds in Figure 8a (upper diagram). The smaller the critical current density, which means, the higher the conductor temperature, the larger is $d\Delta J_{Crit}/d\Delta\zeta$ [$(A/m^2)/m^{-3} s^{-1}$) = A m s], the derivative of the variations of critical current density and relaxation rate, $\zeta$, of the thin film YBaCuO 123 superconductor. During warm-up, this means that the ratio $\Delta J_{Crit}/d\Delta\zeta$ increases with increasing simulation time (or duration of an experiment), as follows from integration of Eq. (3b), i. e. of the curve $d\Delta J_{Crit}/d\Delta\zeta$,

$$\ln(\Delta J_{Crit}) = \int d\Delta\zeta + C \qquad (7a)$$

under the condition $\Delta\zeta(J_{Crit}) = 0$ at $J_{Crit} = J_{Crit0}$ = const (almost zero decay or recombination rates that compensate each other). This yields

$$\Delta J_{Crit} = \Delta J_{Crit0} \exp(-\Delta\zeta) \qquad (7b)$$

Variation of critical current density, coupled by the exponential factor, $\exp(-\Delta\zeta)$ in Eq. (7b), to the decay rate, decreases strongly if the relaxation rate, $\Delta\zeta$, increases. On the other hand, variation of the relaxation rate is very small if temperature is close to the phase transition, compare Figure 13 in the Appendix. If the system continuously creates new, intermediate dynamic equilibrium states, decay rates in this Figure are equal to relaxation rates. This means: Near the phase transition, variations of the decay rate, like a variation $\exp(-20) \leq \Delta\zeta$ [$m^{-3} s^{-1}$] $\leq \exp(-1)$ results in a variation of approximately $2 \cdot 10^{-9} \leq \Delta J_{Crit}/\Delta J_{Crit0} \leq 0.37$. Over the total period, variations of critical current density and of relaxation rates obviously are weakly correlated numerically.

By analogy, conclusions in the same way can be drawn from the derivative $d\Delta J_{Crit}/d\Delta\tau$ of critical current density with relaxation time, τ (the red diamonds in Figure 8a, upper diagram). But because of the plus-sign, the integral of $d\Delta J_{Crit}/d\Delta\tau$ would diverge too strongly to allow conclusions on how critical current density could be increased by variation of relaxation time within reasonable duration of experiments or numerical simulations (the question is how relaxation time can directly be tailored at all).

But finite, possibly tailored *decay rates*, $\Delta\zeta$ [(A/m$^2$)/m$^{-3}$ s$^{-1}$) could be helpful to protect $J_{Crit}$ from decrease under disturbances (large $d\Delta J_{Crit}/d\Delta\zeta$ may "heal" deficiencies). This is because the top and bottom diagrams in Figure 8a explain that, as mentioned, at temperature close to critical temperature, and this is the interval of most interest, the (small) critical current density reacts more sensitive to the (large) variations of $d\Delta J_{Crit}/d\Delta\zeta$. Conversely, at temperature far away from the superconducting/normal conducting phase transition (which means, at large $J_{Crit}$), the variation $d\Delta J_{Crit}/d\Delta\zeta$ is comparatively small,

The above question how to increase $d\Delta J_{Crit}/d\Delta\zeta$ at temperature close to the phase transition has to be answered with respect to solely the *proper* superconductor material.

In type I superconductors, for application of Eq. (1b) with the Ginzburg-Landau exponent n, it is the de-pairing or pair-breaking current density up to which the said correlations may be investigated (this means, as long as the kinetic energy of the super-current carriers does not exceed condensation energy to the superconducting state).

In type II superconductors, it is in addition

(i) the critical current density, at the transition from zero to flux flow resistance (magnetic flux quanta leaving their pinning sites) and

(ii) decoupling by layers or other aggregates of non-super-conducting material between neighbouring superconductor thin films that both limit the range of applicability of Eq. (1b).

No doubt, perfect crystallography and stoichiometry are the first requisites to be fulfilled (a single crystal would be an optimum sample). But when this is fulfilled, what then would be suitable to promote the rates $d\Delta J_{Crit}/d\Delta\zeta$?

The model [11] and its application to calculate relaxation rates within the same range (items i and ii) could give an answer: Since relaxation rate is inversely proportional to relaxation time (compare Figures 12a,b and 13), it is relaxation time that has to be decreased in order to increase $d\Delta J_{Crit}/d\Delta\zeta$. Apart from constraints set by "coefficients of fractional parentage" and the "time of flight" concept, which both are just formalistic issues, it is essentially the density of electron pair states, $n_S$, and its temperature dependence that both enter into the calculations of relaxation time in this model.

Smaller (than usually applied values) $dn_S/dt$, according to the calculation steps in [11], would contribute to reduction of relaxation time, which means future materials development might increase $d\Delta J_{Crit}/d\Delta\zeta$ provided magnitude and temperature dependence of the energy gap in the superconductor could be modified.

Figure 8a, lower diagram, and Figure 8b show that $d\Delta J_{Crit}/d\Delta\zeta$ sensitively depends on the energy gap, $\Delta E(T)$. In the calculations, its temperature dependence is modelled using the expression $\Delta E(t) = 1.73\, \Delta E_0\, (1 - T/T_{Crit})^{0.5}$ for $T/T_{Crit} \rightarrow 1$ (Eq. 2-54 of [4]). The calculations in the upper diagram of Figure 8a apply $\Delta E_0 = 60$ meV. In the bottom diagram, the impact of $\Delta E_0$ on $d\Delta J_{Crit}/d\Delta\zeta$, in the interval $50 \leq \Delta E_0 \leq 70$ meV, is very strong (near $J_{Crit} = 10^8$ A/m$^2$, $d\Delta J_{Crit}/d\Delta\zeta$ increases by about a factor of four when $\Delta E_0$ is increased from 50 to 60 meV).

Accordingly, size of the energy gap is correlated with $d\Delta J_{Crit}/d\Delta\zeta$, still from solely numerically arguments. Are the variables also *causally* ordered? The model [11] explains relaxation time as resulting from a two-step process within two successive time intervals, compare Sects. 3.2.1 and 3.2.2 of this reference. Because of the Pauli exclusion principle (and in view of the Racah coefficients), calculation of total lifetime, τ, has to proceed in a step-wise, sequential manner by the following steps (a) and (b):

(a) Decay widths $\Gamma_{ij}$ for two arbitrary particles i and j, that determine, by correlation, an "intrinsic lifetime", $\tau_1 = (h/2\pi)/\Gamma_{ij}$, of the non-equilibrium state, to be taken for re-combination of particles i and j out of a very large number, N. It requires a tiny, finite time interval needed first to *initiate* the proper condensation or recombination event to a pair. "Correlation", in the sense used in this model, strictly speaking means exchange of "information" between two quantum states; in reality, and from analogy in nuclear physics (with a pion mediating n,p-coupling), it is the exchange of phonons that mediate binding interaction between single particles, like the pion does in the Deuteron.

(b) Contributions $\tau_2 = (h/2\pi)/\Delta E(T)$ resulting from the uncertainty principle (h indicates Planck's constant), for the *proper* condensation or recombination, energy dependent event, once the particles i and j are identified in step (a) (note that this condition specifies the qualitative assumptions made in [13]).

Decreasing size of the energy gap, $\Delta E(T)$, which happens under increase of temperature, increases relaxation time, roughly $\tau = \tau_1 + \tau_2$. Since relaxation time, $\tau$, is inversely proportional to relaxation rate, $\zeta$, decreasing $\Delta E(T)$, under increasing temperature, decreases relaxation rates. As mentioned, seen from the physics behind, the higher the temperature, provided $T < T_{Crit}$, the more decay electrons have to be re-organised to pairs in a new equilibrium by the two steps (a) and (b).

The larger the energy gap, the more will excitations (decay of electron pairs) be blocked, and the more is an entrance channel opened for increased zero-loss current transport. Single (decay) electrons cannot contribute, current transport by single electrons is resistive. Therefore, the larger the energy gap, at given constant temperature, the larger critical current density (red diamonds in Figure 8b).

Accordingly, since the energy gap, $\Delta E(T)$, *physically* (arrow 2 in Figure 8b) correlates with critical current density, both variables, $\Delta E(T)$ and $J_{Crit}(T)$, are definitely (because causally) correlated. This result is of practical use since modifications of $\Delta E_0$ can easier be realized and controlled (IR-absorption, ultra-sonic or tunnelling experiments) than modifications of relaxation rates.

Yet, via arrow 3, the energy gap is causally correlated with relaxation rate, and relaxation rate, via arrow 4, causally correlates with critical current density. Large relaxation rates quickly provide electron pairs to "repair" (support) critical current density. Another uni-directional, causal relation (the reverse of arrow 4, after a 180 degrees turn) is not true.

In summary, arrow 4 indicates, as the final result of this paper, the causally, but *uni-directionally* oriented correlation between relaxation rate and critical current density.

Neutron and heavy ion irradiation are well-known tools to provide effective pinning centres (tracks or point defects). Irradiation increases critical current density, causes variations of irreversibility line and variations of the value of the energy gap. The same can be achieved, again a well-known procedure, with tiny concentrations of paramagnetic impurities (however, gapless superconductivity, yet with strong pair correlations, cannot be explained by the curves in Figure 8a,b).

Another potential concerns the relation between de-pairing critical current density, $J_d$, and coherence length, $\xi$. In high temperature superconductor materials, $\xi$ is very small. Since $J_d \sim 1/\xi$, a correlation could exist, via $J_{Crit} \sim \Delta E$ (at constant temperature), between also relaxation rate and coherence length.

Materials development thus might perhaps be improved if these apparently existing correlations and their potentials could be exploited. Search for new superconductor materials, from the correlations listed above, with tailored energy gap could significantly increase critical current density or, reversely, allow to determine relaxation rates arising from thermal disturbances by measuring critical current density.

## 6     Summary

Near the superconducting/normal conducting phase transition, quench, if any, starts locally. This is the consequence of non-uniform temperature fields, $T(x,y,t)$, that result from disturbances and from inhomogeneity of materials and transport properties. For the same reason, quench neither starts at exactly the same time nor at the same position within of the conductor cross section.

Temperature dependence of $J_{Crit}$ on the exponent n in $J_{Crit}(x,y,t) = J_{Crit}(x,y,t = 0) [1 - T(x,y,t)/T_{Crit}]^n$ is important for calculation of temperature excursions and for prediction of a quench. Convergence tests have confirmed that safely the exponent should apply the Ginzburg-Landau value n = 1.5.

Critical superconductor parameters $J_{Crit}$, $T_{Crit}$ and lower and upper critical fields, $B_{Crit1,2}$, all have to be understood as local values. Local temperature differences, $T(x,y,t) - T_{Crit}(x,y,t)$, and local ratios, $T(x,y,t)/T_{Crit}(x,y,t)$ in modelling critical current density, have to be applied instead of simple (standard) differences $T - T_{Crit0}$ and ratios $T/T_{Crit0}$. This decision is important for success when investigating the said correlations and for stability calculations.

During warm-up, critical temperature exists only as a convergence *limes* of a series of equilibrium superconductor states and corresponding equilibrium temperatures, which all are increasingly distant on the time scale. The *limes* cannot be obtained in standard experiments or simulations if a disturbance at temperature already close to the superconducting/normal conducting phase transition. Critical temperature, if understood as a *sharply, with zero tolerance defined* quantity, accordingly is a fiction.

Relaxation rates, critical current density and the energy gap of superconductors are physically correlated. In particular, arrow 4 in Figure 8b indicates, as the final result of this paper, a causal but uni-directionally oriented correlation between relaxation rate and critical current density. This result might open a guide line for materials development to increase critical current density or, reversely, provide relaxation rates from measured critical current density.

Experiments to verify these predictions would be difficult to perform. But this is not new physics. The results reported in this paper simply reflect aspects of many-particle physics, analogies from nuclear physics and standard methods of multi-component heat transfer in superconductor thin films.

Thank you for your patience. This paper results from my lectures "Applied Superconductivity" given in the Department of Physics at the Unversity of Wuerzburg., Germany.

Comments and nominations of potential referees are highly welcome.

## 7    References

1   Reiss H, Finite element simulation of temperature and current distribution in a superconductor, and a cell model for flux flow resistivity – Interim results, J. Supercond. Nov. Magn. **29** (2016) 1405 – 1422 2   Reiss H, Stability of a (2G) Coated, Thin Film YBaCuO 123 Superconductor - Intermediate Summary, J. Supercond. Nov. Magn. (2020), https://doi.org/10.1007/s10948-020-05590-3

3   Reiss H, An Attempt to Improve Understanding of the Physics behind Superconductor Phase Transitions and Stability, http://arxiv.org/abs/2102.05944 (2021), accepted for publication in Cryogenics (2021)

4   Tinkham M, Introduction to superconductivity, Robert E. Krieger Publ. Co., Malarbar, Florida, reprinted edition (1980)

5   de Gennes P G, Superconductivity of metals and alloys, W. A. Benjamin, Inc., New York (1966)

6   Clem J R, Bumble B, Raider S I, Gallagher W J, Shih Y C, Ambegaokar-Baratoff--Ginzburg-Landau crossover effects on the critical current density of granular superconductors, Phys. Rev. B 35 (1987) 6637 - 6642

7   Djupmyr M, The role of temperature for the critical current density in high-temperature superconductors and heterostructures, Doctoral Thesis, University of Stuttgart, Germany (2008)

8   Janos K., Kus P, Temperature dependence of the critical current density in Y-Ba-Cu-O thin films, Czech. Journal of Physics **40,** 335 – 340 (1990)

9   García-Fornaris I, Planas A., Muné P, Jardim R, Govea-Alcaide E, Temperature Dependence of the Intergranular Critical Current Density in Uniaxially Pressed $Bi_{1.65}Pb_{0.35}Sr_2Ca_2Cu_3O_{10+\delta}$

Samples, Journal of Superconductivity and Novel Magnetism **23** (2010) 1511 – 1516

10  Senatore C, Barth Chr, Bonura M, Kulich M, Mondonico G, Field and temperature scaling of the critical current density in commercial REBCO coated conductors, Superconductor Science and Technology **29** (2016) 014002. 10.1088/0953-2048/29/1/014002

11  Reiss H, A microscopic model of superconductor stability, J. Supercond. Nov. Magn. **26** (2013) 593 - 617

12  Phelan P E, Flik M I, Tien C L, Radiative properties of superconducting Y-Ba-Cu-O thin films, Journal of Heat Transfer, Transacts. ASME 113 (1991) 487 – 493

13  Buckel W, Kleiner R, Superconductivity, Fundamentals and Applications, Wiley-VCH Verlag, 2nd Ed. Transl. of the 6th German Ed. by R. Huebener (2004)

14  Gray K E, Long K R, Adkins C J, Measurements of the liftime of excitations in superconducting aluminium, The Philosophical Magazine **20**\50164\5 (1969) 273 - 278

15  Gray K E, Steady state measurements of the quasiparticle lifetime in superconducting aluminium, J. Phys. F: Metal Phys. (1971, Vol. 1) 290 - 308 16   Annett J, Superconductivity, Superfluids and Condensates, Oxford Master Series in Condensed Matter Physics, Oxford University Press  (2004)

17   Reiss H, Stability considerations using a microscopic stability model applied to a 2G thin film coated superconductor, J. Supercond. Nov. Magn. **31** (2018) 959 - 979

18   Juarez E, Trillaud D, Zermeño V, Grilli F, Advanced electromagnetic modeling of large-scale high temperature superconductor systems based on H and T-A formulations. Superconductor Science and Technology. **34** (2021), 10.1088/1361-6668/abde87

19   Reiss H, The Additive Approximation for Heat Transfer and for Stability Calculations in a Multi-filamentary Superconductor - Part B,  J. Supercond. Nov. Magn. **33** (2020) 629 – 660

20   Reiss H, Radiative Transfer, non-Transparency, stability  against quench in superconductors and their correlations, J. Supercond. Nov.  Magn. **32** (2019) 1529 -1569

21   Reiss H, Inhomogeneous temperature fields, current distribution, stability and heat transfer in superconductor 1G multi-filaments, J. Supercond. Nov. Magn. **29** (2016) 1449 - 1465

**Appendix**

(1) Results obtained from calculation of the stability function,

$$0 \leq \Phi(t) = 1 - \int J_{Crit}[T(x,y,t), B(x,y,t)] \, dA / \int J_{Crit}[T(x,y,t_0), B(x,y,t_0)] \, dA \leq 1 \qquad (8)$$

are shown in Figure 9a,b. The Figure demonstrates the impact of variations $T_{Crit}$ uniform against $T_{Crit} = T_{Crit}(x,y,t)$ and, in more detail, of the exponent n in $J_{Crit}(T) = J_{Crit0} (1 - T/T_{Crit})^n$. The stability function is explained in more detail in [1 - 3].

(2) The Finite Element (FE) to solve Fourier's differential equation in the present, very complicated conductor geometry, is embedded in a comprehensive, "master scheme" (see the flow chart, Figure 11a). It applies a multiple-step method that incorporates combinations of repeated, step by step applications of the proper FE cycles with intermediate, analytic calculations of electrical and magnetic variables.

In few, exceptional cases, local T < 77 K were observed at positions outside the proper superconductor cross sections of the five turns, like in Figure 11c. There a "cold spot" appears within the impregnation material. But the results, within the repetitions of the FE iteration cycles, are adjusted to the minimum temperature, T = 77.0000 K, and iterations could be continued without restarts of the FE program. This correction had no observable impacts on the obtained temperature field T(x,y,t) within the (proper) superconductor thin films.

# Figures

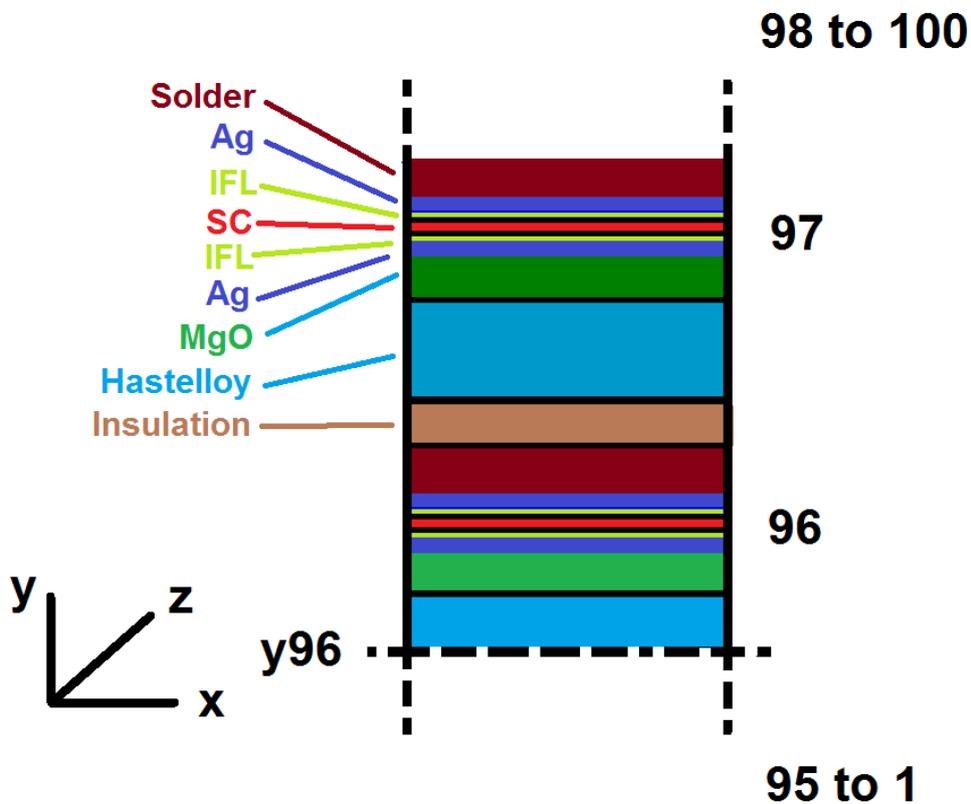

Figure 1 Simulation scheme of coil and of conductor geometry (detail of Figure 1 of [17], here showing turns 96 and 97 of a coil; schematic, not to scale). The coil (in total 100 turns) is prepared from a coated, YBaCuO 123, thin film superconductor. Crystallographic c-axis of the YBaCuO-layers is parallel to y-axis of the co-ordinate system. Conductor architecture and dimensions are standard. Superconductor (SC) layer thickness (red sections) is 2 µm, width 6 mm; thickness and width of the Ag elements (lilac) is the same, width of the interfacial layers (IFL, light green) is 40 nm (the IFL are included to simulate surface roughness and diffusion of species between the SC and its neighbouring Ag- and MgO-layers, respectively). Dimensions of the other conductor components are given in Table 1 of [17]. The thick, dashed-dotted line at the bottom of this diagram indicates the artificial axis of vertical symmetry introduced for support of the Finite Element part of the simulations, compare text for explanation.

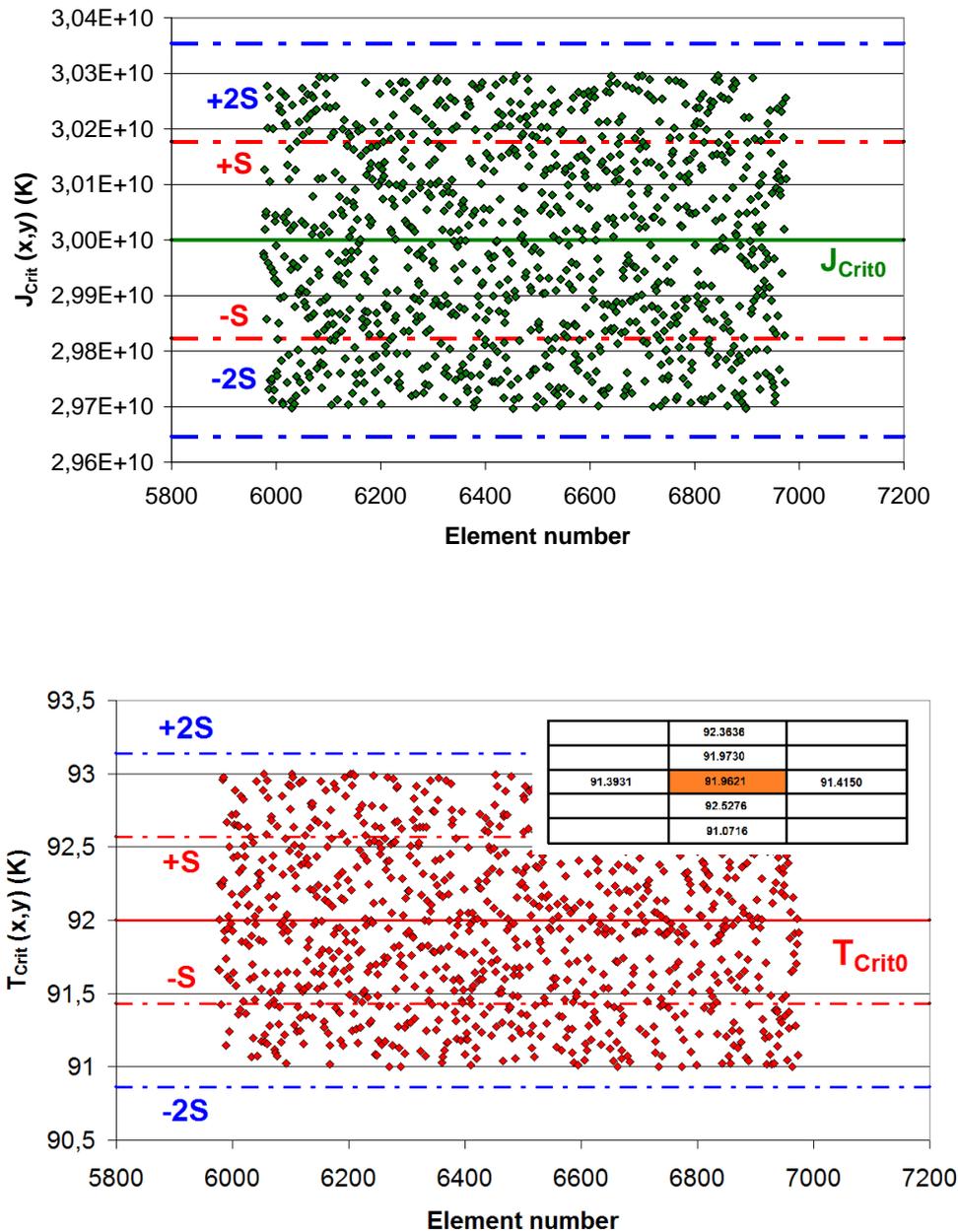

Figure 2 Random variations $\Delta J_{Crit0}$ of critical current density (upper diagram) and of $\Delta T_{Crit0}$ of critical temperature (below). The variations $\Delta J_{Crit0}$ are within 1 percent around the mean (thin film) value $J_{Crit0} = 3 \cdot 10^{10}$ A/m$^2$ of YBaCuO 123 in zero magnetic field and at T = 77 K, the variations $\Delta T_{Crit0}$ are within 1 K around the mean value $T_{Crit0}$ = 92 K in zero magnetic field. In the lower diagram, the inset shows element temperatures in the immediate neighbourhood of the centroid of turn 96 (orange rectangle). Solid green and red lines indicate mean values, blue and red, dashed-dotted lines are mean-square deviations. In both diagrams, random variations and their mean-square deviations are the start point for subsequent, overall variations of these parameters in that multiples of these are used in the simulations (Figure 5b).

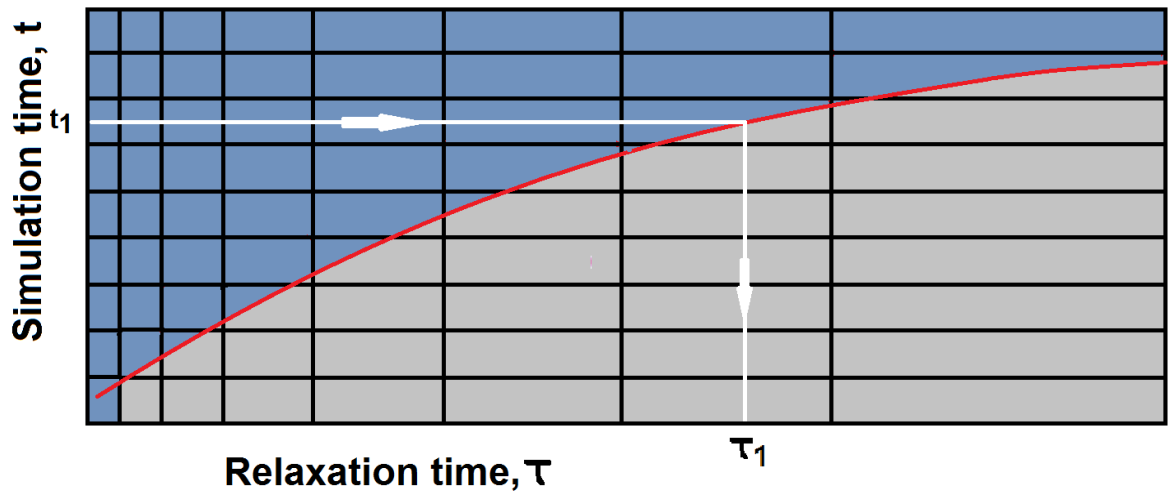

Figure 3 Simulation procedure explaing the excursion of relaxation time (horizontal axis), with simulation time and, correspondingly, temperature (schematic). During warm-up under a disturbance, temperature increases, $T = T(t)$, that generates an increasing number, $n(t)$, of electrons (resulting from decayed electron pairs). Relaxation time, $\tau$, to arrive at a new equilibrium state, thus increases. Length of the horizontal bars thus reflects increase of $\tau$, as a function of simulation time. The red curve illustrates the following chain of dependencies (or origins of events): $dT(t_0)/dt > 0$ (warm up) → $dn(T(t_0)/dt > 0$ (non-equilibrium) → relaxation → $\tau_1$ → $dn_S(T(t_1)/dt = 0$ (equilibrium state reached) → next simulation step $dT(t_2)/dt > 0$, and to be continued.

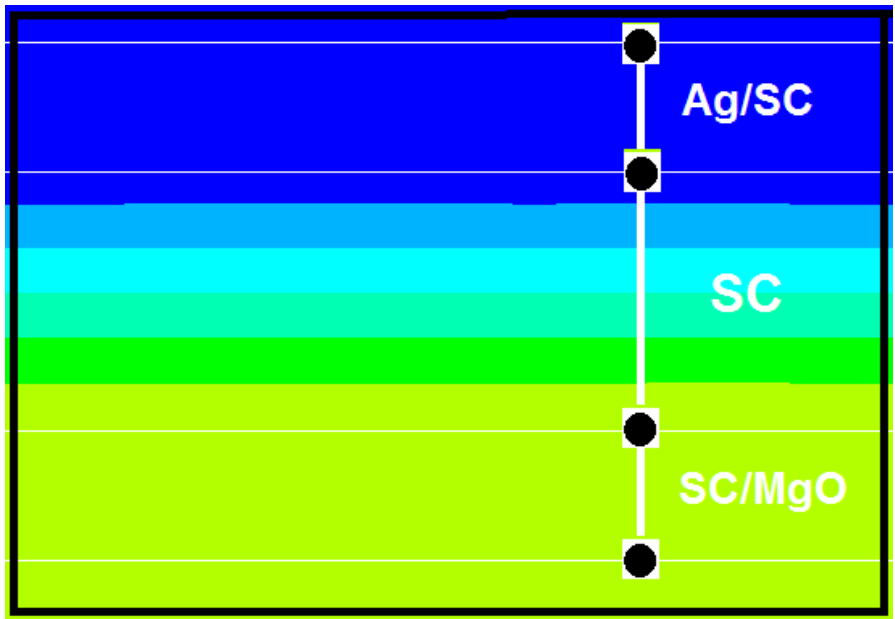

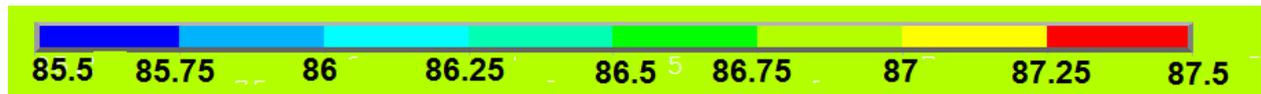

Figure 4a  Temperature distribution (strongly magnified detail) in the 2 µm, thin film superconductor (SC) and its interfacial layers in turn 96 when convergence is achieved. Results are shown near the right end of the conductor. The approximately constant temperature gradient within the SC layer (six nearly equally spaced intervals) demonstrates linear increase of temperature at t = 4.1 ms simulation time. Thickness of the interfacial layers Ag/SC and SC/MgO is 1 µm. Non-zero thickness of these layers serves for modelling surfaces roughness in the Finite Element part of the simulations using plane, 4-node (not shell) elements.

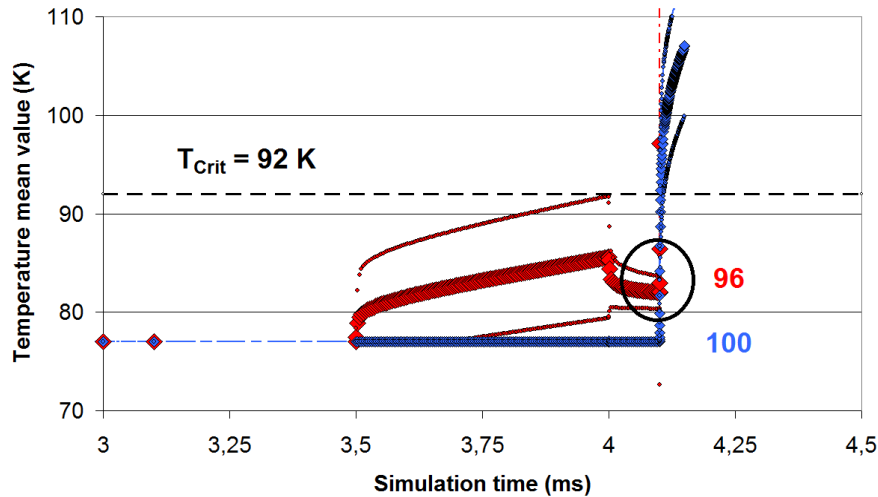

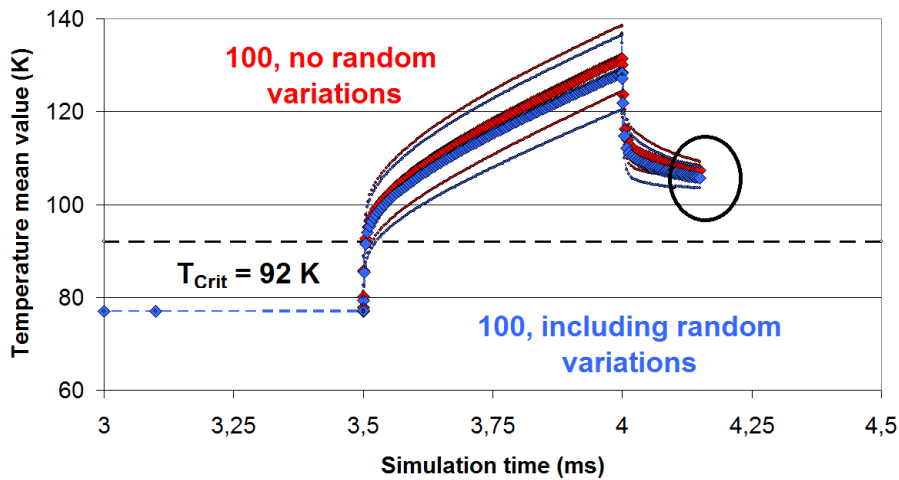

Figure 4b Mean element temperature (solid diamonds) in turns 96 and 100, with an apparent temperature run-away beginning at t = 3.5 ms and the previously explained (compare [1 - 3]) "saw-tooth" convergence behaviour of T(x,y,t). The large black "convergence circles" identify converged conductor temperature when it is finally obtained under the multiple iterations (see the explanations to Figure 11a,b in the Appendix). Thin solid circles denote standard deviations. The simulations apply the anisotropy ratio, $D_{ab}/D_c = 10$, of the thermal diffusivity, D, in crystallographic ab-plane and c-axis direction of the superconductor. In the expression for the critical current density, $J_{Crit}(x,y,t) = J_{Crit}(x,y,t_0) [1 - T(x,y,t)/T_{Crit}]^n$, the exponent n is 1.5. The Figure shows a first temperature run-away (a local quench).

Upper diagram: Results calculated *without fault current.* Calculations include random variations of $T_{Crit}$, $J_{Crit}$ and $B_{Crit1,2}$. Because of increasing divergences, the simulations are stopped at t > 4.15 ms shortly after start of the second local quench.

Lower diagram: Results obtained *with inclusion of the fault current.* In the outer turns (here turn 100), almost no difference within the standard deviations can be detected if calculations are performed with or without the random variations of $T_{Crit}$, $J_{Crit}$ and $B_{Crit1,2}$.

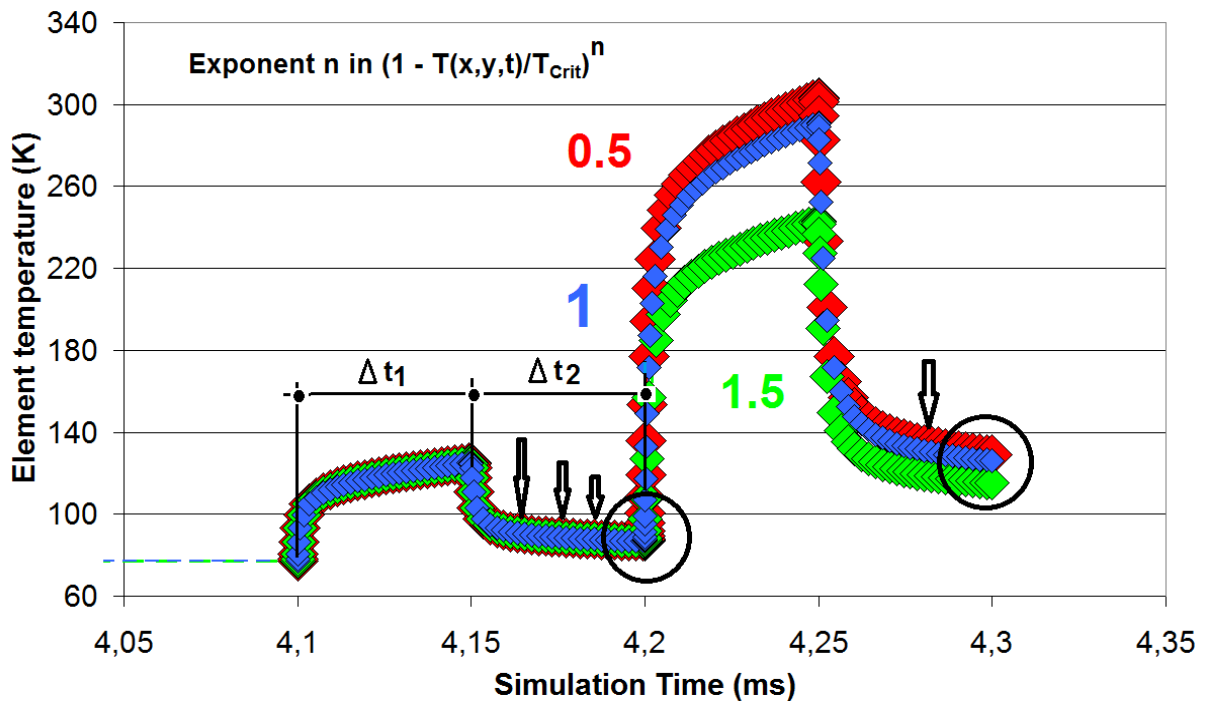

Figure 5a Superconductor element temperature (centroid of turn 96) of the YBaCuO 123 thin film superconductor vs. simulation time. The Figure explains temperature excursion and its convergence in Figure 5a,b of [17] and Figure 13b (part 2 and 3) of [2]. The arrows indicate results obtained during the iterations (sub-steps, i) within a particular load step, j, before convergence is finally achieved (these are the black convergence circles). The saw-tooth behaviour of conductor temperature beginning at t = 4.1 and 4.2 ms is shown for different values of the exponent n in the relation for the YBaCuO 123 thin film superconductor $J_{Crit}(x,y,t) = J_{Crit}(x,y,t_0) [1 - T(x,y,t)/T_{Crit}]^n$, for constant anisotropy ratio, $D_{ab}/D_c = 5$, of the thermal diffusivity, D, $I_{Transp}/I_{Crit} = 1$ (no fault, just nominal transport current). Results are calculated with the random variations $\Delta J_{Crit0}$, $\Delta T_{Crit0}$ and $\Delta B_{Crit20}$ (Figure 2) of the critical parameters. That the local temperature run-away at t = 4.1 ms (and also at t = 4.2 ms) in reality predicts a total quench becomes clear in Figure 5b, lower diagram. Differences resulting from the exponents n are significant (but the stability function, trivially, is not affected since $T > T_{Crit}$).

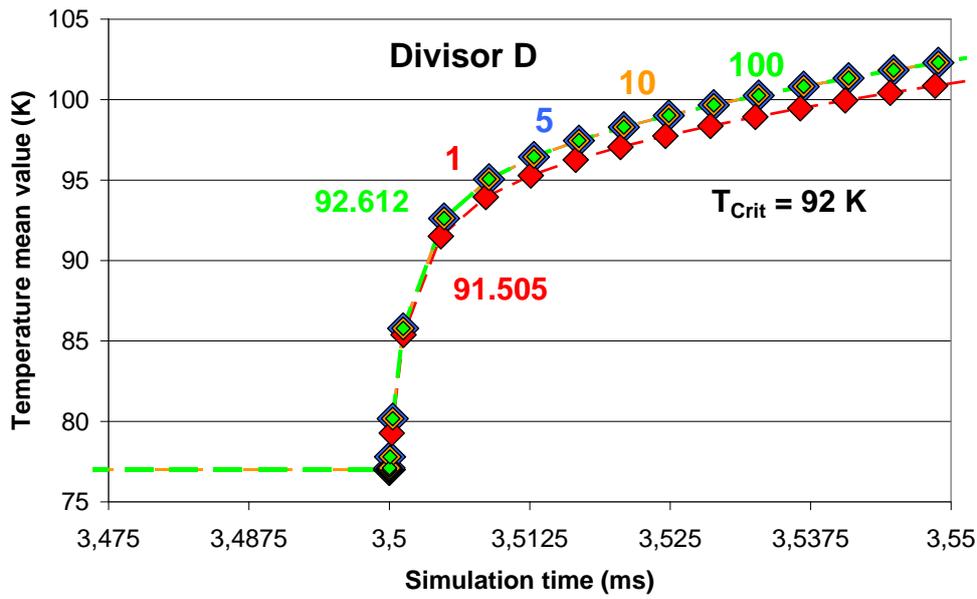

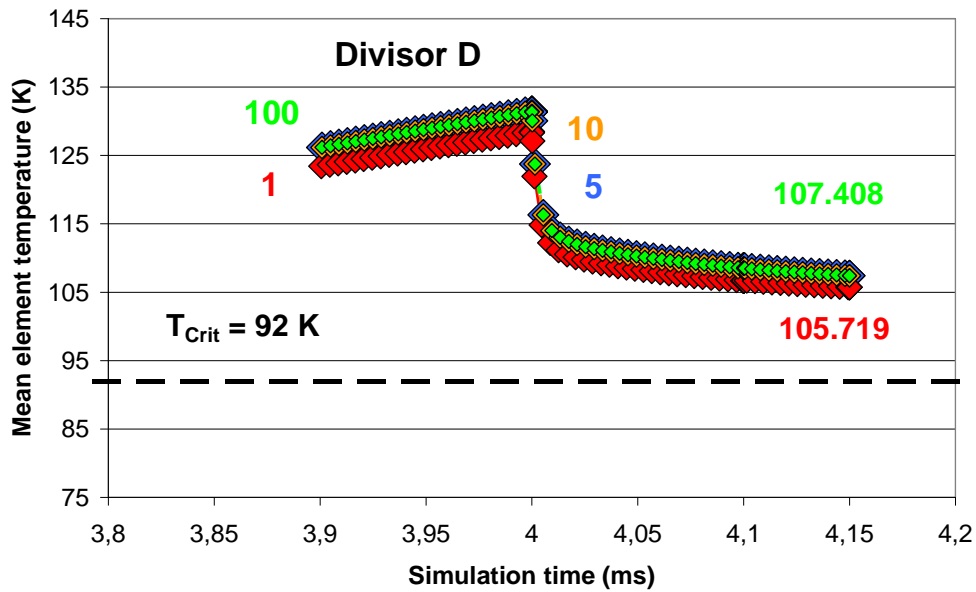

Figure 5b Impact initiated by variations $\Delta T_{Crit0}$, $\Delta J_{Crit0}$ and $\Delta B_{Crit20}$ on mean element temperature in turn 100; results are shown at the beginning of the iterations (upper diagram) and within the convergence circle (below). To demonstrate the impact, the values $\Delta T_{Crit0}$ and $\Delta J_{Crit0}$ of Figure 2 and the $\Delta B_{Crit20}$ are divided by a divisor $1 \leq D \leq 100$. Temperature $T(x,y,t)$ increases the larger the divisor (the smaller the uncertainty $\Delta T_{Crit0}$, $\Delta J_{Crit0}$ and $\Delta B_{Crit20}$). Near $T_{Crit}$ (above), the total effect on the mean of $T(x,y,t)$ in turn 100 amounts to about 1.1 K, in the convergence circle (diagram below) it increases to about 1.7 K. In turn 96, the effect tends to larger values, but this is subject to the boundary condition to turns 95 and below.

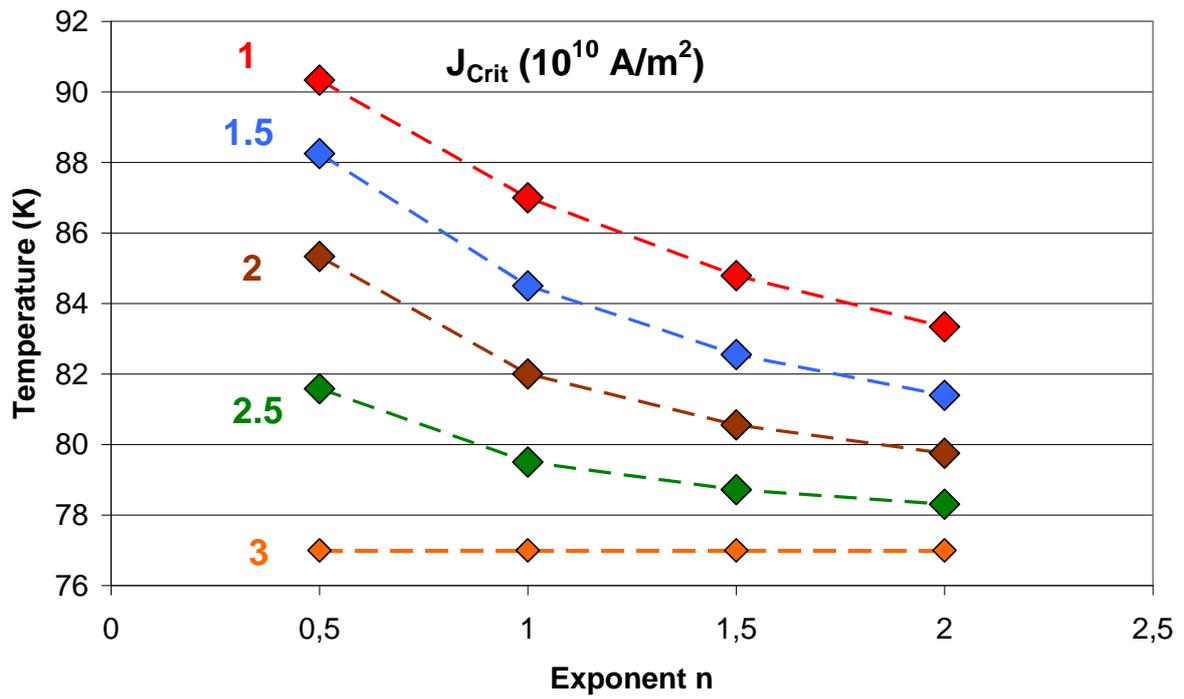

Figure 6 Dependency of conductor temperature on the exponent n in $J_{Crit}(x,y,t) = J_{Crit0}(x,y,t)[1 - T(x,y,t)/T_{Crit}]^n$. The Figure yields conductor temperature, T, if Eq. (1b) is solved for given $J_{Crit}$ and given exponent n, which yields Eq. (4).

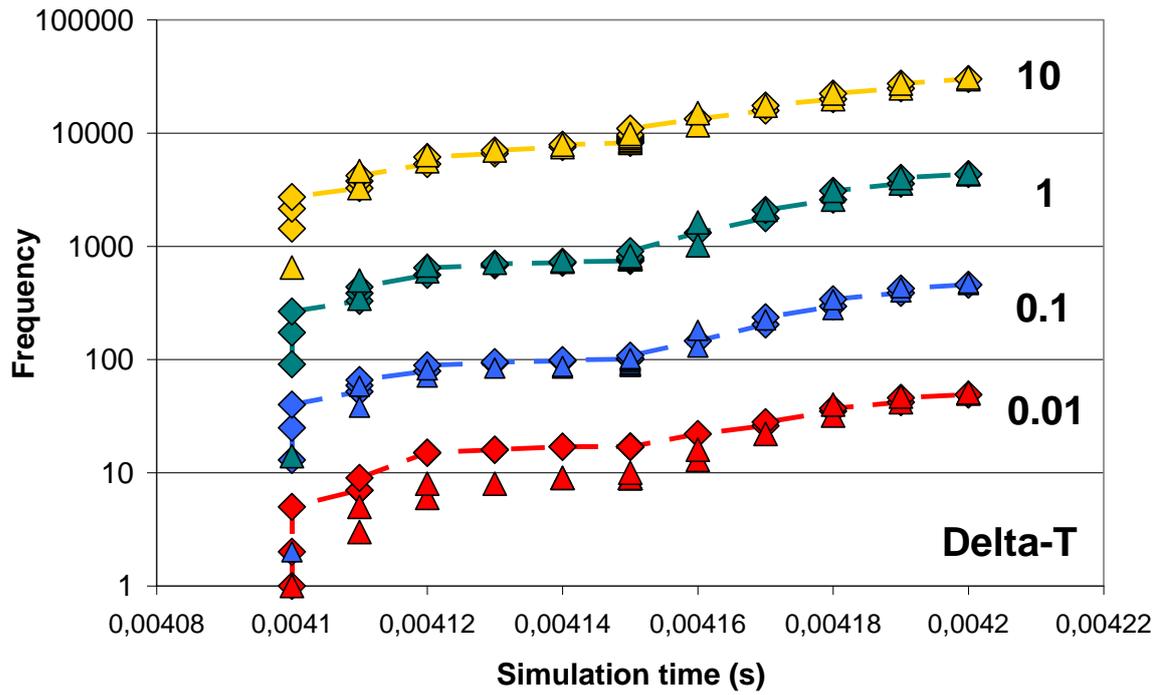

Figure 7 Statistical frequency distribution obtained under from either [T(x,y,t) - $T_{Crit,0}$] < δT (diamonds) or from [T(x,y,t) - $T_{Crit}$(x,y,t)] < δT (triangles) within limits (constants) δT = 0.01, 0.1, 1 and 10 K. Results are calculated from the temperature distributions in turn 96 for the anisotropy ratio, $D_{ab}/D_c$ = 5, and for the exponent n = 1.5 in $J_{Crit}$(x,y,t) = $J_{Crit}$(x,y,t = 0) [1 - T(x,y,t)/$T_{Crit}$]$^n$. No fault, just nominal current, and with $I_{Transp}/I_{Crit}$ = 1.

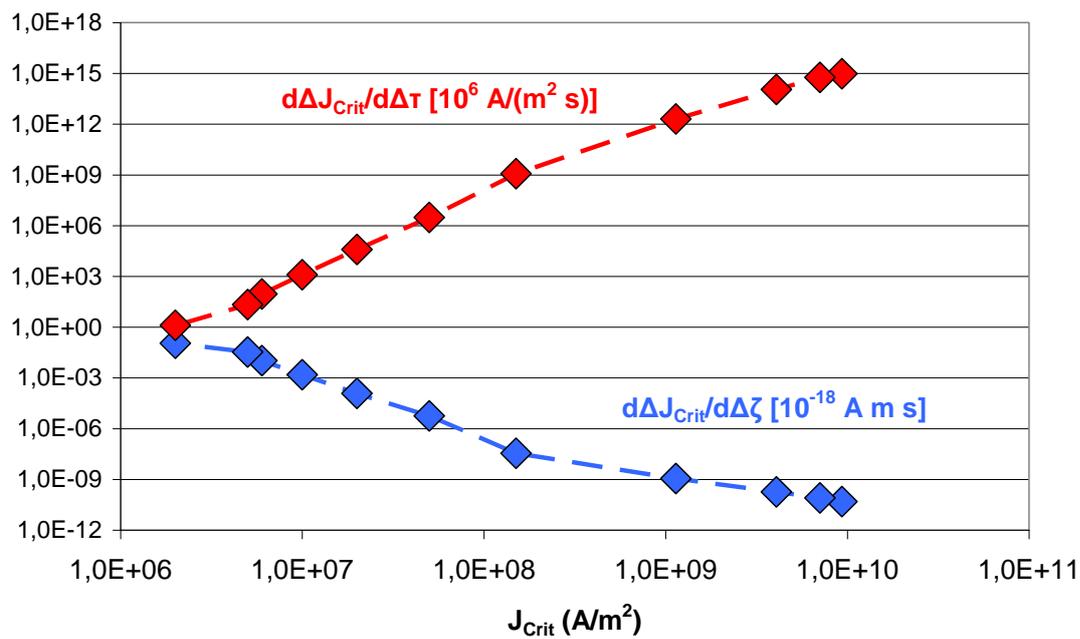

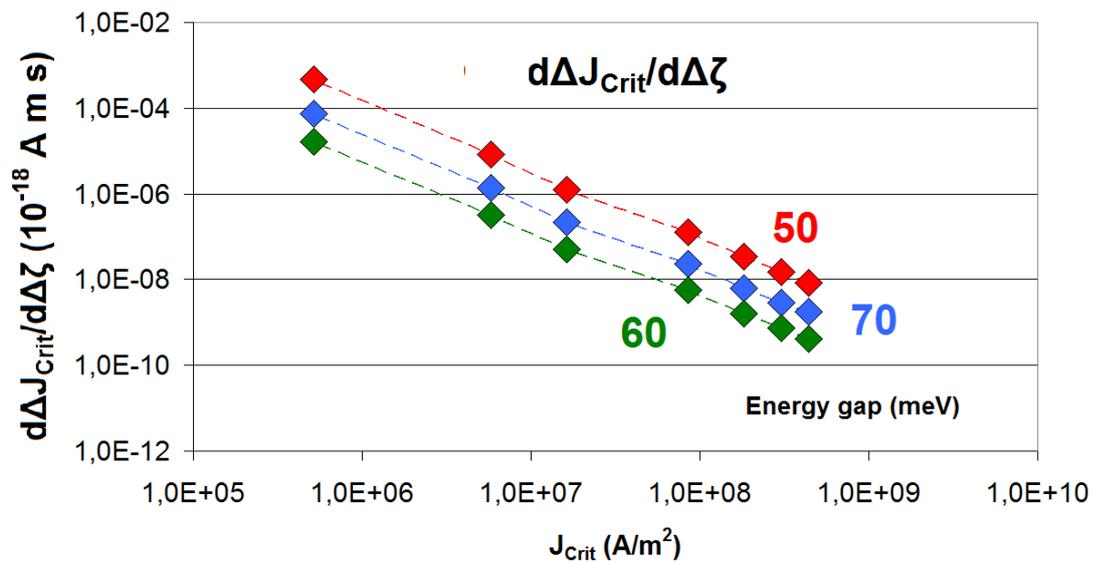

Figure 8a Upper diagram: Differential critical current density, $J_{Crit}$, per relaxation time, $\tau$ and per relaxation rate, $\zeta$, calculated with $\Delta E_0 = 60$ meV. Below: $d\Delta J_{Crit}/d\Delta\zeta$ calculated with different $\Delta E_0$; in both diagrams, results are plotted vs. $J_{Crit}$.

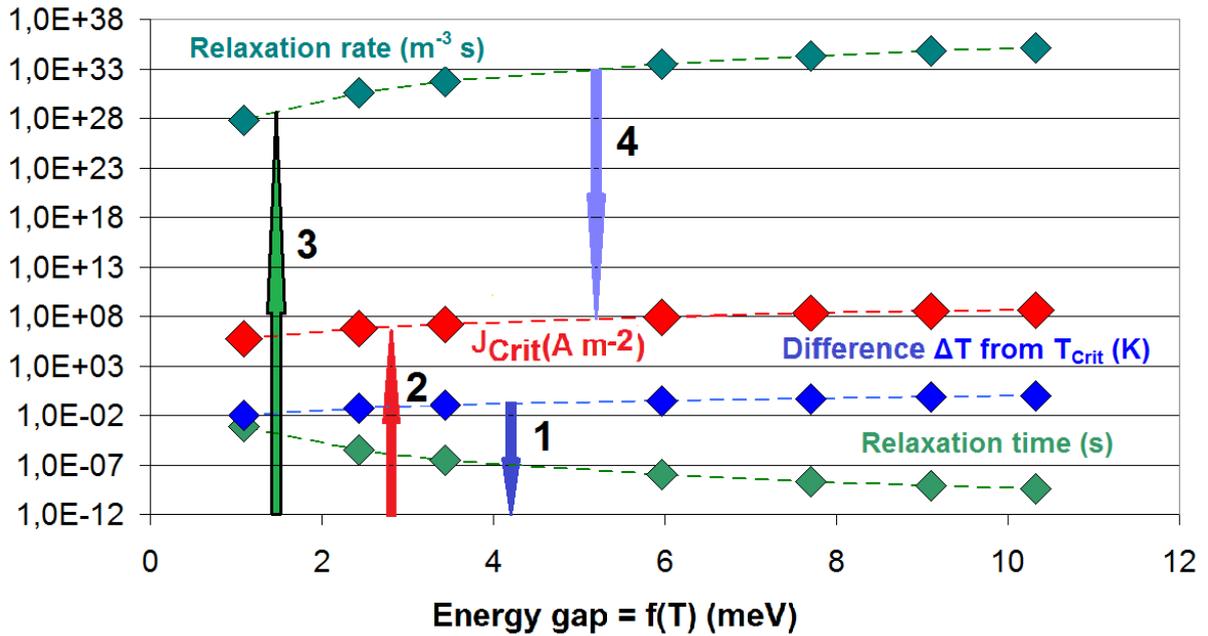

Figure 8b Relaxation time τ, relaxation rate ζ, difference ΔT = $T_{Crit}$ - T(t) and $J_{Crit}$ vs. energy gap ΔE(T) near critical temperature. The τ- and ζ-curves are calculated from [11] and $J_{Crit}$ from Eq. (1b) using n = 3/2. Arrows indicate quantitative correlations and causal relations (the latter are uni-directional) 1: temperature (here the difference ΔT) → energy gap (causal), 2: energy gap → $J_{Crit}$ (causal), 3 (in parallel to 2): energy gap → relaxation rate (causal), 4: relaxation rate → $J_{Crit}$. Accordingly, arrow 4 indicates, as the final result of this paper, the causally but uni-directionally oriented correlation between relaxation rate and critical current density.

# Figures in the Appendix

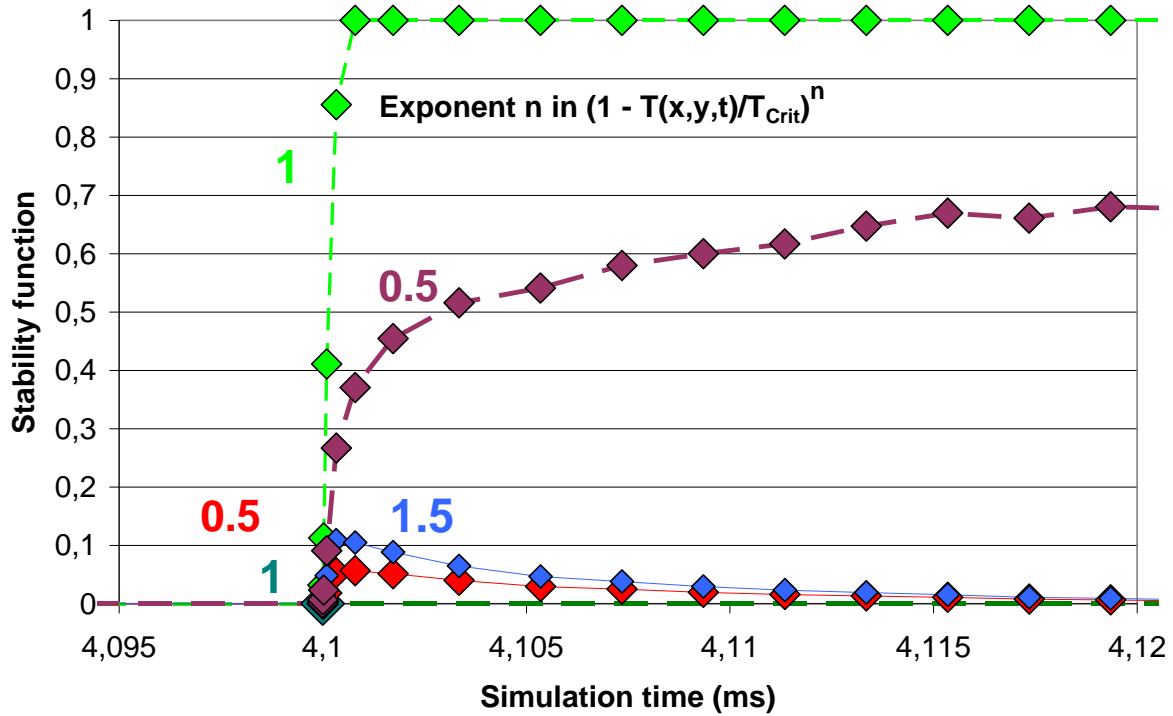

Figure 9a Stability function, Φ(t), of the YBaCuO 123 thin film superconductor (detail near t = 4.1 ms) taking into account elements of turn 96 (dark-brown diamonds). By its steep increase of Φ(t) at t = 4.1 ms, the Figure clearly identifies the onset of a first quench. No fault current, just nominal transport current, $I_{Transp}/I_{Crit}$ = 1. The calculations apply different values n of the exponent in the relation $J_{Crit}(x,y,t)$ = $J_{Crit0}(x,y,t_0)$ $[1 - T(x,y,t)/T_{Crit}]^n$. In this Figure, critical current density $J_{Crit0}$ is considered as *uniform*, $J_{Crit0}$ = 3 $10^{10}$ A/m$^2$ at 77 K in all elements, but of course, $J_{Crit}(x,y,t) = J_{Crit0}$ $(1 - T/T_{Crit})^n$, and *without* the other statistical fluctuations. Resistive current limiting is provided, apparently almost completely, by turn 96 (light-green and dark-brown diamonds). At t > 4.115 ms, zero loss transport current would decrease by about 30 percent when instead of n = 1 (light-green) the value of the exponent n is reduced to 0.5 (dark-brown diamonds).

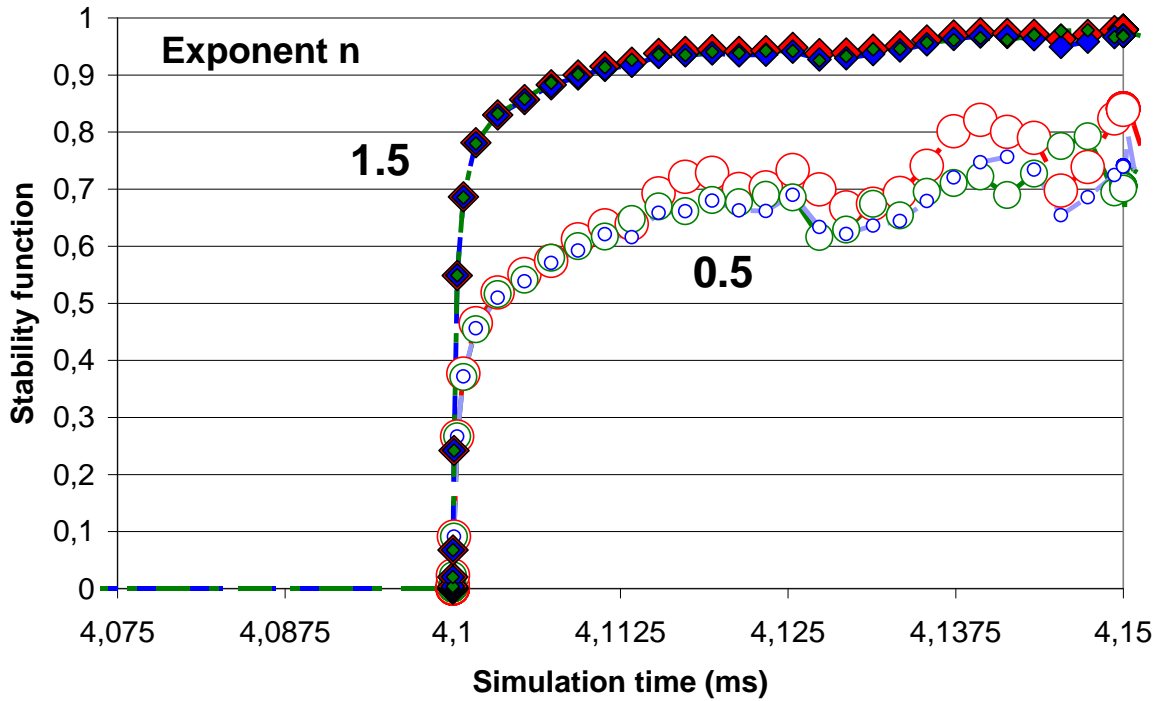

Figure 9b Stability function, $\Phi(t)$, of the YBaCuO 123 thin film superconductor (detail near t = 4.1 ms) of the YBaCuO 123 thin film superconductor (all elements of turn 96). The Figure compares results obtained with application of $T_{Crit} = T_{Crit0}$ uniform and $T_{Crit} = T_{Crit}(x,y,t)$. No fault, just nominal transport current. Like in Figure 9a, onset of the quench is clearly identified from the sudden increase of $\Phi(t)$ at t = 4.1 ms. Results are obtained with values n = 1.5 (solid diamonds) and 0.5 (open circles) of the exponent in the relation $J_{Crit}(x,y,t) = J_{Crit}(x,y,t) [1 - T(x,y,t)/T_{Crit}]^n$, for constant ratio $D_{ab}/D_c = 5$ of the thermal diffusivity and for $I_{Transp}/I_{Crit} = 1$. Critical current density $J_{Crit0}$ is *uniform*, $J_{Crit0} = 3\ 10^{10}$ A/m$^2$ at 77 K in all elements. For the decision whether the superconductor is in zero loss, flux flow or Ohmic states, the calculation compares $T(x,y,t)$ with locally different values $T_{Crit}(x,y,t)$ under variations within the maximum spacing $\Delta T_{Crit0}$ (the values in Figure 2). Red, blue and green symbols denote results obtained with differences and ratios (i) $T(x,y,t) - T_{Crit0}$ and $T(x,y,t)/T_{Crit0}$, (ii) $T(x,y,t) - T_{Crit}(x,y,t)$ and $T(x,y,t)/T_{Crit0}$ and (iii), $T(x,y,t) - T_{Crit}(x,y,t)$ and $T(x,y,t)/T_{Crit}(x,y,t)$, respectively; the differences in $\Phi(t)$ are greater if n = 0.5. At values of $\Phi > 0.7$, zero loss transport current would decrease by about 20 percent when instead of n = 1.5 the value n = 0.5 is applied. Differences among results obtained with options (i) to (iii) at t > 4.14 ms (open circles), though below 15 percent, cannot be neglected, which means item (iii) has to be considered for reliable stability calculations.

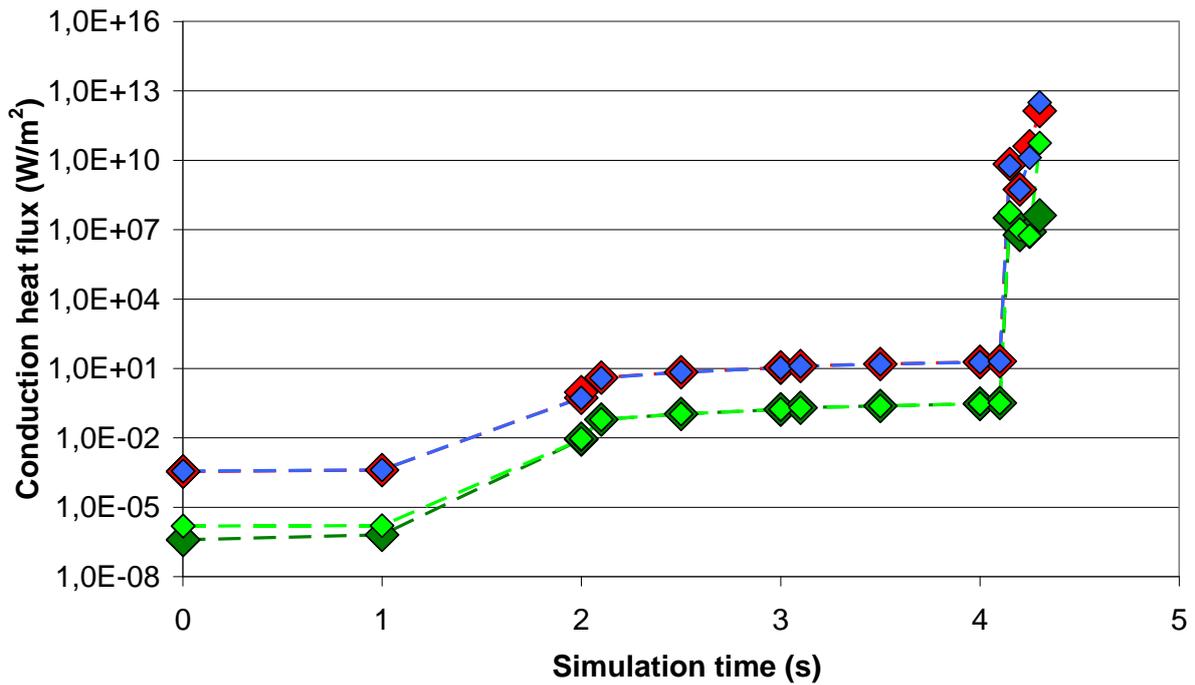

Figure 9c Conduction heat flux at the interface between the superconducting centroid element to neighbouring films in turn 96. Solid red and blue diamonds indicate vertical, light and dark red-green diamonds horizontal heat flux. Results have been obtained with the exponent n = 1.5 in $J_{Crit}(x,y,t) = J_{Crit}(x,y,t) [1 - T(x,y,t)/T_{Crit}]^n$, for constant ratio $D_{ab}/D_c = 5$ of the thermal diffusivity and for $I_{Transp}/I_{Crit} = 1$.

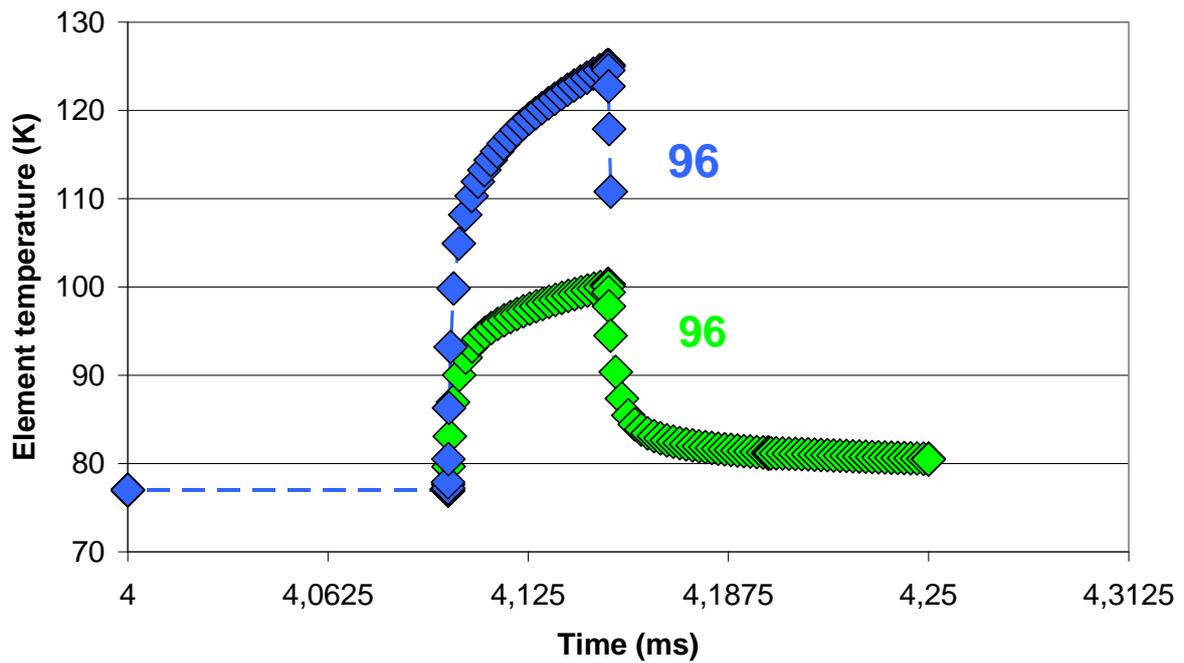

Figure 10 Impact on centroid element temperature in turn 96 if thermal conductivity, $\lambda_{ab}(T)$ and $\lambda_c(T)$, in the crystallographic ab-plane and c-axis direction of the superconductor in turn 98 are randomly scattered (light-green diamonds). Scattering is by ± 10 percent around the temperature dependent, mean value. Anisotropy of $\lambda_{ab}(T)$ and $\lambda_c(T)$ is 10. The blue diamonds are copied from Figure 5a.

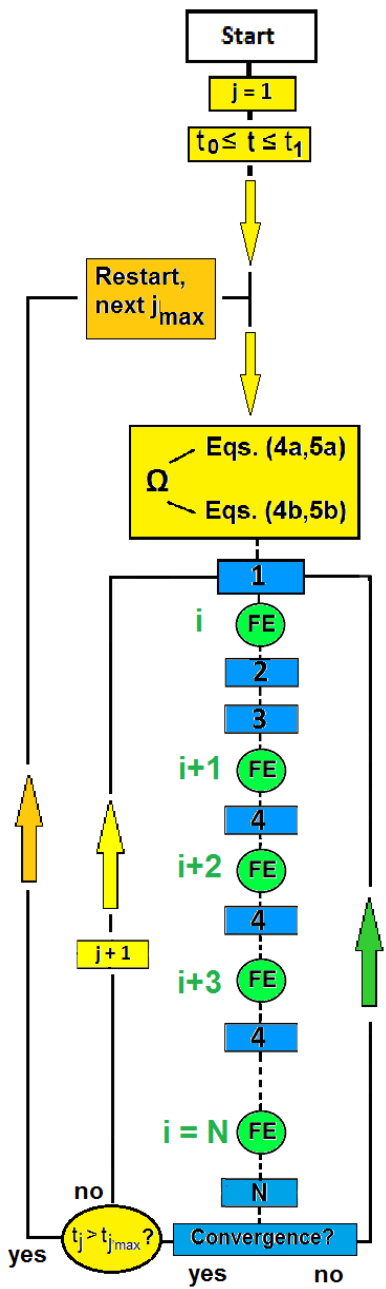

Figure 11a Flow chart (the previously introduced "master" scheme) showing two iteration cycles (i, j) and one time loop ($t_j$) of the numerical simulation: <u>Light-green circles and indices, i</u>: Sub-steps, the proper Finite Element (FE) calculations; <u>Light-yellow indices, j</u>: Load steps involving FE and, within the blue rectangles, critical current, magnetic field and resistance (flux flow, Ohmic) calculations; <u>Dark-yellow indices, t</u>: Time loop, lines of a matrix **M** (equation numbers shown in the rectangles) reflect their definition in Section 6 of [3]. The blue rectangles with sub-step numbers i = 1, 2, 3,...N are defined as **1:** First FE step, j, with data input of start values of temperature distribution, specific resistances, critical parameters of J, B and of initial (uniform) transport current distribution or of single, isolated radiation heat pulses, respectively; **2:** Results obtained after the first FE step (i), if converged, for the same parameters in the *same* load-step, j; calculation of $T_{Crit}$, $B_{Crit}$, $J_{Crit}$; **3:** Calculation of resistance network and of transport current distribution (if applicable), all to be used as data input into the next FE calculation (sub-step i + 1), within the *same* load step, j; **4:** Results like in **2**; Sub-steps **5, 6,...N:** Results like in **3** or **4;** convergence yes or no ? If "no", return to **1** (iteration i = 1, in the same load step, j). If "yes" go to next load step j + 1, continue with **1**. The number N of FE cycles (green circles) might strongly increase computation time. Length of simulation time, $t \leq t_{max}$, within each of the individual intervals, with $t_{max}$ indicating the maximum time of a corresponding particular interval, is selected according to the different transit times, source functions, different radiation propagation mechanisms, different ratios of solid conduction and radiation, and to different wavelengths. By the time-loop, t, Figure 11a is an extension of Figure 12 of [19]. Like Figure 11a,b, Figures 12a,b and 13 (below) are shown here solely for convenience of the readers. By citations of the originals, they do not constitute self-plagiarism and, by the references made to [3] http://arxiv.org/abs/2102.05944 (2021), do not violate transfer of copy-right assigned to the paper JCRY103325 submitted to Cryogenics (still in the production stage).

*Figure reprinted by permission from Springer Nature and Copyright Clearance Center under License No 519 117 151 0983 from Journal of Superconductivity and Novel Magnetism (Aug 23, 2019): Harald Reiss, The Additive Approximation for Heat Transfer and for Stability Calculations in a Multi-Filamentary Superconductor - Part B (Figure 12).*

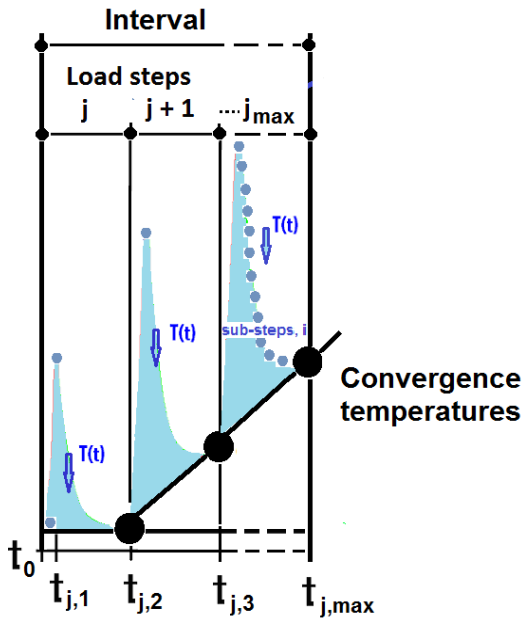

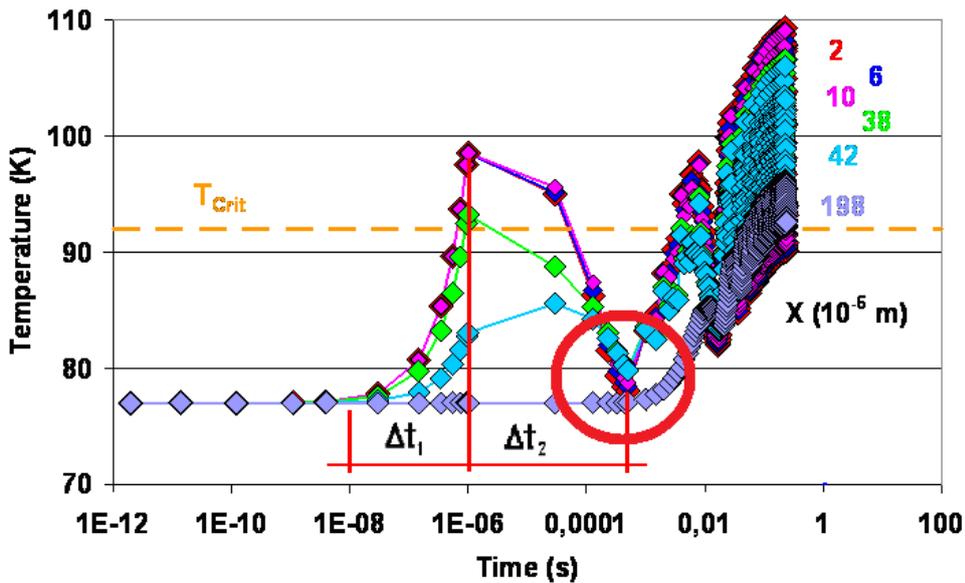

Figure 11b Solution scheme (above) used in the numerical simulations including the Finite Element (FE) procedure integrated in the master scheme (Figure 11a, schematic, not to scale; compare text). Copied from Figures 14b,c of [3], the upper diagram shows the "saw-tooth" behaviour of conductor temperature during the iterations and the convergence circle. For incident radiation pulses, a Monte Carlo simulation is performed in the interval $\Delta t_1$. Solution of Fourier's differential equation, to calculate excursion with time of conductor temperature, T(t) proceeds in the interval $\Delta t_2$; we have $\Delta t_1 << \Delta t_1$. The dashed blue curve (above) schematically indicates conductor temperature, T(t), or any temperature-dependent parameter like specific resistance or specific heat or thermal conductivity and may also indicate results of the analytical calculations (like $J_{Crit}$ or the stability function) in the master scheme. Convergence temperature (or, accordingly, convergence of the temperature-dependent parameters) is indicated by the large, solid black circles, at the end of each of the load steps, j. The diagram below, as an example, shows convergence of local element temperatures after an isolated but periodic thermal excitation by heat pulses incident on a YBaCuO 123 filament. Results refer to positions X from the centre of the circular filament. Convergence behaviour of the temperature resembles the saw tooth-like behaviour. The large red circle indicates convergence temperature (the convergence cycles in Figures 4b and 5a,b of the present paper) after a number $N_{lt}$ (a...k) of iterations (compare Figure 11a) within the load steps, j (the load steps in this diagram are identified by computation time). At time t > 0.01 s, the saw tooth-like convergence behaviour can no longer be resolved, because of the logarithmic plot.

*Figure reprinted by permission from Springer Nature and Copyright Clearance Center under License No 519 187 106 7998 from Journal of Superconductivity and Novel Magnetism (Aug 23, 2019): Harald Reiss, The Additive Approximation for Heat Transfer and for Stability Calculations in a Multi-Filamentary Superconductor - Part B (Figure 14).*

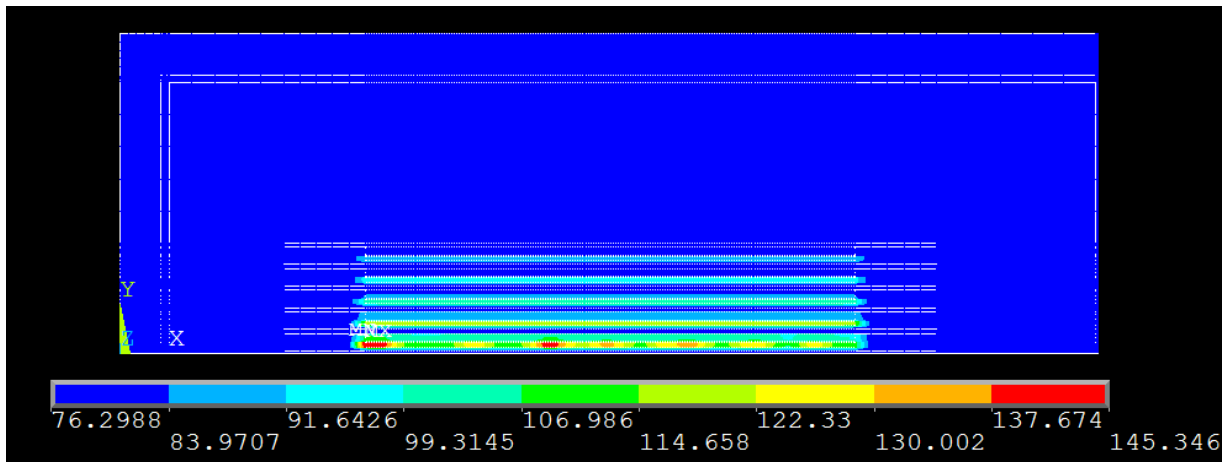

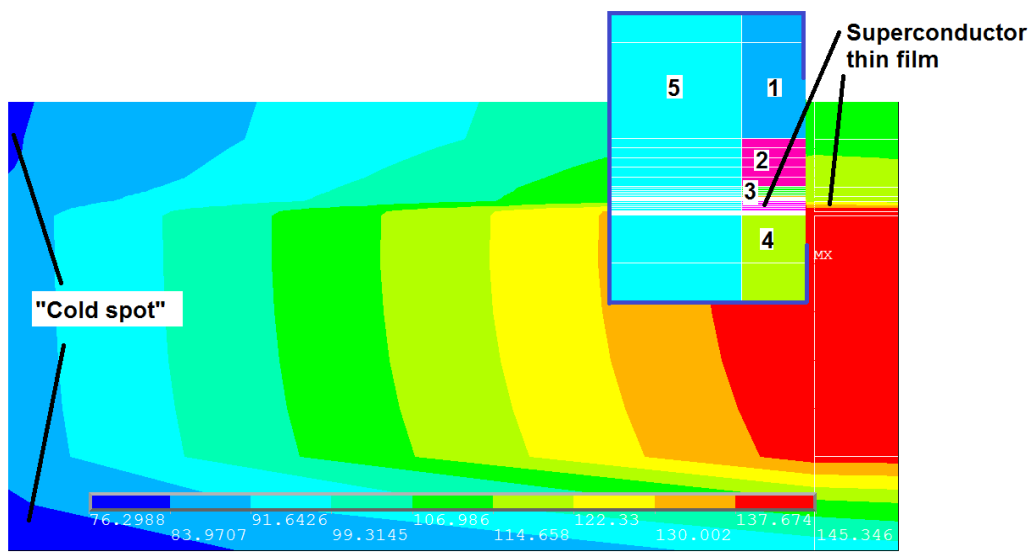

Figure 11c Nodal temperature distribution between turns 96 and 100 (above). Temperature, here shown at t = 4.27 ms, has increased against t = 4.2 ms. No fault, just nominal transport current, $I_{Transp}/I_{Crit}$ = 1. Below, the strongly magnified section shows temperature distribution at the left end of the superconductor thin film in turn 96 with a "Cold spot" arising during the iterations. The inset in this Figure identifies materials and their positions within the conductor cross section around turn 96: 1 Stabilizer Cu, 2 PbSnAg-solder, 3 metallization Ag, followed by interfacial layer Ag/SC, superconductor (SC) thin film, interfacial layer SC/Buffer layer, 4 buffer layer MgO, 5 Hastelloy. The results indicate increasing convergence problems: The lower temperature limit decreases to 76.2988 instead of 77.0000 K, but this cold spot (here the two, dark-blue regions) is located outside the proper superconductor thin film cross sections. Within the repeated FE iteration cycles, temperature of the cold spot is adjusted to the exact 77.000 K without interruption and restart of the FE simulation.

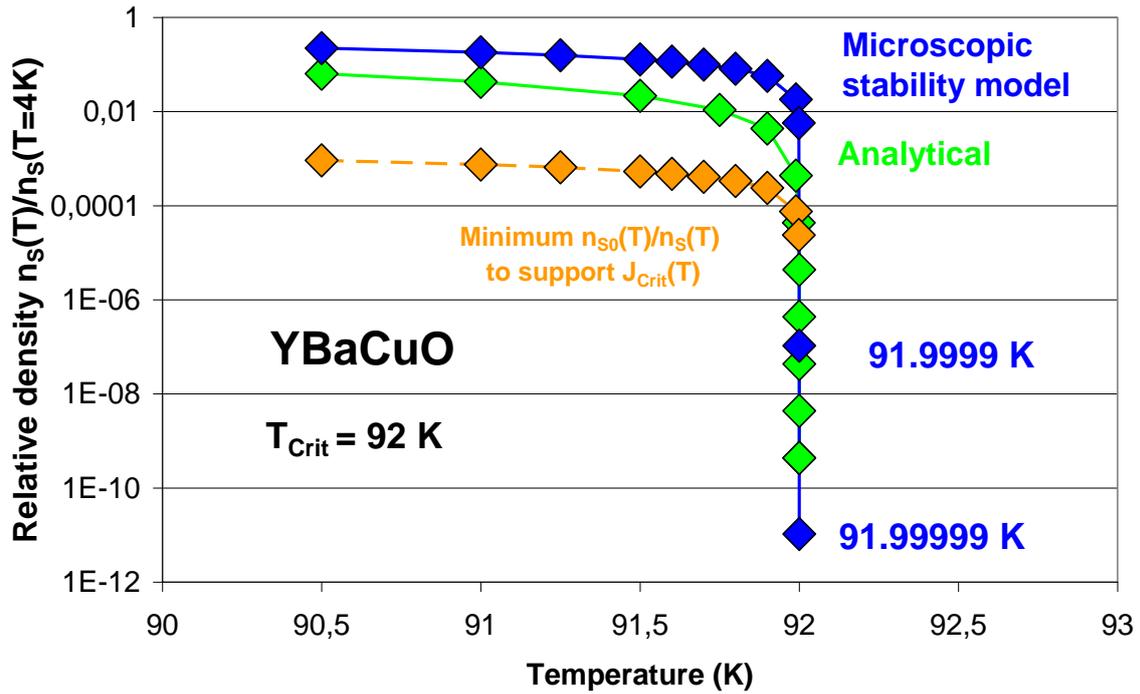

Figure 12a Relative density (the order parameter), $f_S = n_S(T)/n_S(T=4K)$, in dependence of temperature, calculated for the thin film, YBaCuO 123 superconductor (dark-blue diamonds). Dark-yellow diamonds indicate the minimum relative density of electron pairs that would be necessary to generate a critical current density of $3 \cdot 10^{10}$ A/m$^2$ (YBaCuO) at 77 K, in zero magnetic field. The diagram compares predictions of the microscopic stability model with analytical results (light-green) calculated from Eq. (8) in [12]. The Figure is copied from Figure 1a of [3] (is re-plotted for convenience of the reader).

*Figure reprinted by permission from Springer Nature and Copyright Clearance Center under License No 519 187 135 5046 from Journal of Superconductivity and Novel Magnetism (Aug 23, 2019): Harald Reiss, The Additive Approximation for Heat Transfer and for Stability Calculations in a Multi-Filamentary Superconductor - Part B (Figure 8a).*

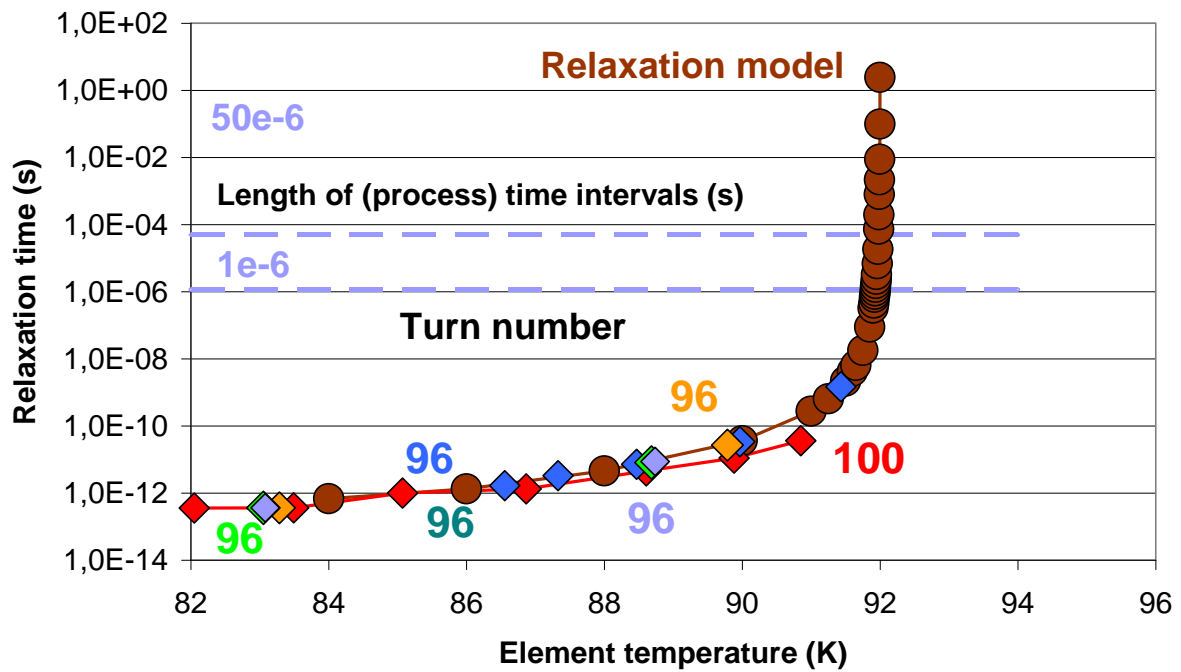

Figure 12b Relaxation time of the superconductor electron system after a disturbance (here the sudden increase of transport current in the windings to a large multiple of the nominal value); results are given vs. element temperature. All solid, dark-brown, circles in this Figure are recalculated and extended against the curve (dark-green diamonds) in Figure 1b of [3] to element temperatures still closer approaching to $T_{Crit}$. Computation time becomes enormous. For just the uppermost data point, at 91.999 K, on a standard, 4-core PC with 3.4 GHz and 16 GB workspace, the calculation under Windows 7 took more than two days (running without any interruption). The Figure is re-plotted for convenience of the reader.

*Figure reprinted by permission from Springer Nature and Copyright Clearance Center under License No 519 188 008 6792 from Journal of Superconductivity and Novel Magnetism (Aug 23, 2019): Harald Reiss, The Additive Approximation for Heat Transfer and for Stability Calculations in a Multi-Filamentary Superconductor - Part B (Figure 11).*

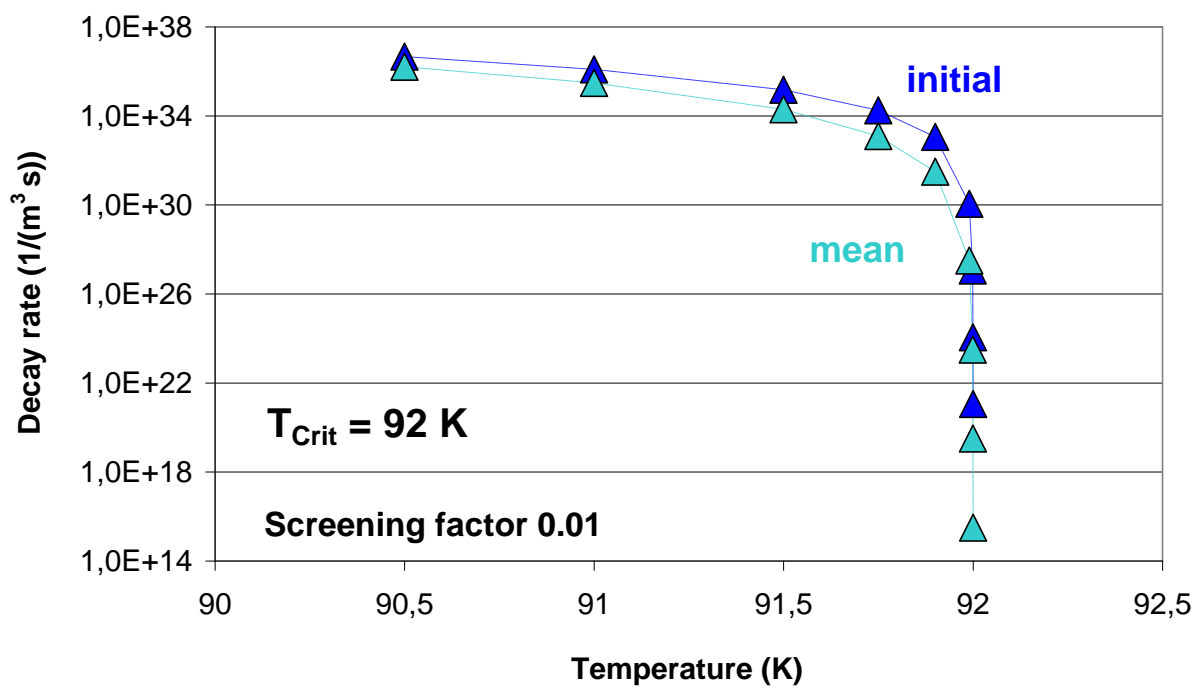

Figure 13 Initial and mean decay or relaxation rates (per unit volume) of thermally excited electron states calculated using a screening factor, $\chi = 0.01$, to the Coulomb potential, in a virtual conductor volume, $V_C$, of the superconductor YBaCuO 123 (see [11] for detailed explanations). If the system continuously creates new (intermediate), dynamic equilibrium states, decay rates in this Figure are equal to relaxation rates. The area below the curves corresponds to the ordered phase (electrons condensed to electron pairs) that is separated, from the thermally disordered phase (electrons from decayed electron pairs), by the classical critical, finite temperature boundary (here the dashed dark-blue and light-green lines). The Figure is copied from Figure 3b of [11].